\documentclass[%
 reprint,
superscriptaddress,
 amsmath,amssymb,
 aps,
pra,
]{revtex4-2}

\usepackage{graphicx}
\usepackage{dcolumn}
\usepackage{bm}
\usepackage{siunitx}

\begin{document}

\preprint{APS/123-QED}

\title{Formation of motile cell clusters in heterogeneous model tumors:\\the role of cell-cell alignment}

\author{Quirine J.~S. Braat}
\affiliation{Department of Applied Physics and Science Education, Eindhoven University of Technology, Eindhoven, The Netherlands}
 
\author{Cornelis Storm}%
\affiliation{Department of Applied Physics and Science Education, Eindhoven University of Technology, Eindhoven, The Netherlands}
\affiliation{Institute for Complex Molecular Systems, Eindhoven University of Technology, Eindhoven, The Netherlands}


\author{Liesbeth M.~C. Janssen}
\email{l.m.c.janssen@tue.nl}
\affiliation{Department of Applied Physics and Science Education, Eindhoven University of Technology, Eindhoven, The Netherlands}
\affiliation{Institute for Complex Molecular Systems, Eindhoven University of Technology, Eindhoven, The Netherlands}

\date{\today}

\begin{abstract}
Circulating tumor cell clusters play an important role in the metastatic cascade. These clusters can acquire a migratory and more invasive phenotype, and coordinate their motion to migrate as a collective. Before such clusters can form by collectively detaching from a primary tumor, however, the cluster must first aggregate in the tumor interior. The mechanism of this cluster formation process is still poorly understood. One of the possible ways for cells to cluster is by aligning their direction of motion with their neighboring cells. This work aims to investigate the role of this cell-cell alignment interaction on the formation of motile cell clusters inside the bulk of a tumor using computer simulations. We employ a Cellular Potts model in which we model a two-dimensional heterogeneous confluent layer containing both motile and non-motile cells. Our results indicate that the degree of clustering is governed by two distinct processes: the formation of clusters due to the presence of cell-cell alignment interactions among motile cells, and the suppression of clustering due to the presence of the dynamic cellular environment (comprised of the non-motile cells). We find that the largest motile clusters are formed for intermediate alignment strengths, contrary to what is observed for motile cells in free space (that is, unimpeded by a dense cellular environment), in which case stronger cell-cell alignment always leads to larger clustering. Our findings suggest that the presence of a densely-packed cellular environment and strong cell-cell alignment inhibits the formation of large migratory clusters within the primary tumor, providing physical insight into potential factors at play during the early stages of metastasis. 
\end{abstract}

\maketitle

\section{Introduction}
Cancer metastasis is the process by which cancer cells spread from a primary tumor site to distant organs or tissues in the body. This process is responsible for the majority of cancer-related deaths \cite{Dillekas_deathsduetometastasis}. The way in which cancer metastasizes depends on a large number of factors, including cell-intrinsic properties such as the invasive potential, and extrinsic properties such as the extracellular environment \cite{Friedl_plasticitycellmigration, Kang_novelPhaseDiagram_tumorInvasion}. There is growing evidence that aggressive metastasis is governed by groups of tumor cells rather than individual tumor cells, thus implying that collective cluster migration is the more effective and perhaps even dominating mechanism  \cite{Giuliano_CTC_VillagetoMetastasize, Chemi_EarlyDissemination_of_CTCs, Aceta_CTCsareprecursorofmetastasis, Friedl_MigrationOfCoordinatedCellClusters}. As cancerous cells detach from the primary tumor and start to migrate through the body, they are considered to be circulating tumor cells (CTCs) or CTC clusters \cite{Cheung_perspectiveonclustermetastasis, Pinearo_relevanteofCTCclustersinbreastcancer}; Typically, most CTC clusters consist of 2 to 20 cells, are relatively small compared to the size of the tumor, and have a relatively high metastatic potential \cite{Aceta_CTCsareprecursorofmetastasis,Matrinex_CTCs_in_zebrafish,Bocci_BiophysicalModel_uncovers_CSD_acrossCancer}. Understanding how these clusters form and migrate is essential for developing new strategies to prevent metastasis from occurring. 

One of the current hypotheses is that cluster formation starts at the primary tumor \cite{Aceta_CTCsareprecursorofmetastasis, Cheung_polyclonalbreastcancer_from_collectivedissemination} via a change in cell phenotype, such as via the epithelial-to-mesenchymal transition (EMT). During the EMT, epithelial cells acquire a more mesenchymal phenotype which results in a loss of cell-cell adhesion or basal-apical orientation, and enhanced migratory and invasive properties \cite{Kalluri_basicsofEMT, Nieto_EMT2016}. This transition plays a role in processes such as embryonic development and wound healing, but, more crucially, a dysregulated EMT can promote cancer metastasis \cite{Kalluri_basicsofEMT}. A full or partial EMT can already manifest itself in the primary tumor to generate a heterogeneous tumor tissue \cite{Bonnomet_invivo_EMT_in_CTCsbreastcancer, Carey_heterogeneous_tumor_exp}, and can lead to hybrid EMT states where cells possess both epithelial and mesenchymal characteristics. Research has shown that such hybrid states can further enable the migration of CTC clusters during various steps of the metastatic cascade \cite{Jolly_ImplicationsOfHybridEMT, Nieto_EMT2016, Bocci_BiophysicalModel_uncovers_CSD_acrossCancer}. The migratory properties of the mesenchymal cells are of particular relevance for cancer progression, as they drive the active migration of cells and could initiate the formation of migrating CTC clusters in the primary tumor. 

If cells start to metastasize as clusters from the primary tumor, there must be a process that causes and stabilizes the collective migration of these cells. From a physical perspective, there are various mechanisms that can lead to collective cell migration \cite{Camley_PhysicalModels_CollectiveCellMotility, Alert_PhysicalModels_of_CollectiveMigration} and therefore also give rise to cluster formation. Examples of cellular clustering mechanisms include differences in cell-cell adhesion \cite{Lv_collectivemigration_in_epithelial_cancerousmonolayer,Nakajima_kinetics_CPM_revised, Beatrici_meanclusterapproach_cellsorting}, differences in velocities \cite{Beatrici_meanclusterapproach_cellsorting, Hallou_tumorheterogeneitypromotescollectiveinvasion}, self-alignment (or velocity alignment) \cite{Kabla_collectivecellmigration}, and cell-cell neighboring alignment \cite{Belmonte_SPPmodelforcellsorting, MartinGomez_CollectiveMotion_PolarAlignment, laang2018coordinated}. Among these, the latter is a distinctive two-body (cell-cell) interaction that, in addition to single-cell properties, can affect the collective behavior in non-trivial ways. Experimentally it is difficult to determine how neighboring alignment interactions govern cellular clustering, but computer simulations afford a systematic means to investigate the effect of these cell-cell interactions on the emergent clustering dynamics. The neighboring alignment mechanism has already shown good agreement with experimental observations in other biological phenomena \cite{Sepulveda_collectivecellmotion, Rappel_selforganizedvortexstate, laang2018coordinated}. However, to the best of our knowledge, the effect of cell-cell alignment on the clustering of cells in a confluent layer has not been studied yet. 

The aim of this work is to identify the effect of cell-cell alignment interactions on (CTC) cluster formation within a model heterogeneous primary tumor. The clusters studied in this work develop inside the primary tumor, i.e.\ before cell detachment starts to play a role. Specifically, we focus on the clustering dynamics of motile cells that reside in a densely-packed tumor environment with many non-motile cells. We use the terms motile and non-motile for the mesenchymal and epithelial cells, respectively, to indicate that we focus on the difference in cell motility between these two cell phenotypes. For simplicity, we model the bulk of the tumor as a two-dimensional (2D) confluent layer, in which the non-motile cells act as a dynamic environment for the migrating motile cells. As a first approximation to the actual continuous distribution between non-motile and highly motile phenotypes, we consider a binary system with cells that are either completely non-motile or motile. We perform simulations using the Cellular Potts Model (CPM), which is an efficient computational framework to study both collective and single cell dynamics in biologically relevant settings \cite{rubenstein2008role, Maree_CPM_book_Hogeweg, Guisoni_ModellingActiveCellMovement, Scianna_ECM, Katsuyoshi_CellAlignment_PolarizedAdhesion, devanny2023signatures}. Indeed, the CPM has already been successfully used in the past to study e.g.\ invasion of migratory cells into a dynamic non-motile cellular background in the absence of neighboring aligning interactions \cite{Kabla_collectivecellmigration, Hallou_tumorheterogeneitypromotescollectiveinvasion}. To account for cell-cell neighboring alignment among the motile cells, we draw inspiration from the physics of active matter and introduce a modified Vicsek alignment mechanism \cite{Vicsek_NovelTypePhaseTransition, Debets_EnhancedPersistence}. This allows us to study clustering for both strong and weakly aligning cells, thereby covering a wide range of possible cell-cell interaction strengths. In addition to cell-cell interactions, we also vary the persistence time to account for the persistent motion of individual motile cells \cite{Li_persistentrandommotion_absenceofsignals}. We show that clustering is strongly affected by both the neighboring alignment interactions and the cellular environment through which these cells have to move. Moreover, our results indicate that tuning neighboring interactions has a different effect on clustering than changing the cell-intrinsic persistent motion, even though both relate to the global orientation of motion of the cells. These simulation results improve our understanding of how clusters may form at the onset of cancer metastasis, i.e., in the bulk of a primary tumor. 

The remainder of the paper is organized as follows. First, we introduce the Cellular Potts Model, describe the cell-cell alignment mechanism, and discuss the characterization of the clustering. Next, we show how the clustering of migrating cells is affected by the cell-cell alignment interactions and the dense environment. For that, we include the analysis of simulations in free space and in a confluent layer both with and without alignment interactions. Finally, we discuss the interplay between the different effects that simultaneously play a role when including both alignment and the cellular environment in the simulations.

\section{Materials and methods}

\subsection{Computational model}
To study how migrating clusters form in a densely-packed heterogeneous environment, we employ the CPM as implemented in CompuCell3D  \cite{Swat_CC3D}. The CPM has originally been formulated by Graner and Glazier to study cell sorting based on differences in cell-cell adhesion \cite{Graner_CellSorting, Glazier_Simulation_of_DAdrivenrearrangement}, but is now widely used to simulate cells and tissue for a broad range of biological phenomena \cite{Maree_CPM_book_Hogeweg, Hirashima_CPMreview_morphogenesis, scianna2013_book,Szabo_review_CPM_tumor}. Briefly, the model represents cells as pixels on a discrete lattice, and the cell dynamics evolves according to the Monte Carlo algorithm \cite{Frenkel_ChapterMonteCarlo}. For each trial move, a random pixel (with cell number~$\sigma_i$) is selected and one of its neighboring pixels (with cell number~$\sigma_j$) is targeted. To determine whether an pixel-copy attempt changes the target site, the change in the Hamiltonian associated with the pixel-copy is calculated. The probability of copying $\sigma_i \rightarrow \sigma_j$ is determined by the Metropolis algorithm \cite{Metropolis_MCMetropolis}:
\begin{equation}
    P(\sigma_i \rightarrow \sigma_j) = 
    \begin{cases}
                1                                       & \Delta \mathcal{H} \leq 0\\
    \exp\left(-\frac{\Delta \mathcal{H}}{T_m}\right)    & \Delta \mathcal{H} > 0
\end{cases}
\end{equation}
where $\Delta \mathcal{H}$ is the change in the Hamiltonian due to the pixel-copy and $T_m$ is an effective temperature that sets the cell-membrane fluctuations \cite{Swat_CC3D}. The timescale of the dynamics is set by a Monte Carlo step (mcs). 

Our simulations account for heterogeneity of the primary tumor by considering two cell types, namely motile and non-motile cells. These two cell types capture the differences in intrinsic motility of mesenchymal and epithelial cell phenotypes observed in biological tumors. The motile cells are driven by an active force that allows for directed cell migration, while the non-motile cells only undergo shape fluctuations. Therefore, we also refer to these motile and non-motile cells as being active and passive, respectively. The non-motile part of the Hamiltonian $\mathcal{H}_0$, i.e.\ without the active force, for both the motile and non-motile cells is given by  \cite{Graner_CellSorting, Glazier_Simulation_of_DAdrivenrearrangement}
\begin{eqnarray}
\mathcal{H}_0 &&  = \mathcal{H}_{adh} + \mathcal{H}_{vol} \label{eq:original_hamiltonian}\\ && = \sum_{\substack{i,j\\    neighbors}} J_{\alpha_i, \alpha_j}(1-\delta(\sigma_i,\sigma_j) ) +\sum_\sigma  \lambda_{V} (V_\sigma-V_t)^2,  \nonumber
\end{eqnarray}

where $\sigma_i$ is the cell number (at pixel $i$) and $\alpha_i$ the corresponding cell type. The cell-cell adhesion term is proportional to $J_{\alpha_i, \alpha_j}$, which depends on the cell types of the adjacent pixels. The Kronecker delta ($\delta(\sigma_i,\sigma_j)$) ensures that a cell does not experience adhesion interactions with itself. The target volume of each cell is given by $V_t$ and the strength of the volume constraint is $\lambda_{V}$. For the motile cells, we include an additional active force contribution that accounts for cell motility. The total change in the Hamiltonian associated with a pixel-copy attempt thus becomes \cite{Guisoni_ModellingActiveCellMovement}
\begin{equation}
    \Delta \mathcal{H} = \Delta \mathcal{H}_0 - \sum_{\substack{\sigma}}
    \kappa \hat{p}_\sigma \cdot \Delta \vec{R}.
    \label{eq:hamiltonian_term_active}
\end{equation}
Here $\kappa$ represents the magnitude of the active force, which is set to $0$ for the passive cells. The unit vector $\hat{p}_\sigma \equiv (\cos(\theta_\sigma), \sin(\theta_\sigma))$ represents the direction of the active force for a given cell $\sigma$, and $\Delta \vec{R}$ is the vector associated with the center-of-mass displacement of the cell during a proposed pixel-copy. When the motile cell is moving in the direction of the active force, the active force term decreases the total energy in the system, thereby promoting motion in this direction. 

To account for cell-cell alignment among the motile cells, we incorporate an alignment mechanism based on the Vicsek model \cite{Vicsek_NovelTypePhaseTransition}. Briefly, this model describes the movement of particles that align their direction of motion with nearby particles in the presence of white noise. Although there are various possible mechanisms by which cells can induce such coordinated motion, the Vicsek model offers a computationally efficient means to mimic an effective alignment interaction between cells. Here, we use an adaptation of the original model where the strength of the alignment interaction can be varied, as in Ref.~\cite{Debets_EnhancedPersistence}. This adapted model allows us to study a wide range of effective alignment interactions and investigate how it changes the emerging dynamics of cells. 

Our adapted Vicsek model is governed by an effective alignment strength and the persistence time characterizing the persistent motion of cells. The evolution of the cell's preferred direction of motion is determined by 
\begin{equation}
    \theta_{\sigma}(t+\Delta t) = \arg\left(\hat{p}_\sigma(t) + \sum_{\sigma'} \gamma \hat{p}_{\sigma'}(t)\right) + \sqrt{\frac{2}{\tau}} \Gamma(\Delta t). 
    \label{eq:angular_direction_both_pa_noise}
\end{equation}
The term in large parentheses models the alignment interaction between the motile cells, where $\gamma$ is a measure for the alignment strength between cell $\sigma$ and the neighboring cells $\sigma'$. We assume that only the motile cells align their motion with their active neighbors. The alignment interaction is only included between motile cells connecting to each other with at least one pair of pixels. Since the alignment interaction is only included for motile cells, the non-motile cells are not considered to have an explicit role in the (re)alignment of the active cell's polarity vector. However, the presence of non-motile cells will affect the cell dynamics indirectly. The last term of Eq.~\ref{eq:angular_direction_both_pa_noise} models the rotational diffusion of a cell's polarization. Here, $\tau$ is the persistence time and $\Gamma(\Delta t)$ is a Gaussian white noise term with zero mean and variance $\Delta t$, i.e., $\langle \Gamma(t)\rangle = 0,\,\langle \Gamma(t) \Gamma(t')\rangle = \delta(t-t')$. Note that $\tau$ is a single-cell property, while $\gamma$ governs the alignment interaction between two neighboring cells. 

The alignment strength $\gamma$ can be varied between $0$ and $1$ to study how aligning cell-cell interactions affect the clustering. When setting $\gamma = 0$, cells migrate independently of other cells and are essentially acting as Active Brownian Particles (ABPs). On the other hand, when $\gamma = 1$, the model reduces to a standard Vicsek formulation where a cell's direction is the average over itself and its direct neighbors. By tuning the alignment strength from $0$ to $1$ we thus cover a large range of effective cell-cell interactions. 

\subsection{Simulation details}

\begin{table}[b!]
\centering
\caption{The full set of simulation parameters. The parameters that are explicitly varied in this work are indicated in bold. \label{tab:simparams}} 
\begin{tabular}{|c|>{\centering}p{4.5cm}|c|}
\hline
\textbf{Symbol} & \textbf{Definition} & \textbf{Value} \\\hline
$(x, y)$ & Box-size dimensions & (400, 400) pixels \\ \hline 
$t_{max}$ & Maximum simulation time & 100 000 mcs \\ \hline 
$T_m$ & Effective temperature & $10.0$ \\ \hline
$J_{PP}$ & Adhesion strength between passive cells & $10.0$ \\ \hline
$J_{AP}$ & Adhesion strength between active and passive cells & $10.0$ \\ \hline
$J_{AA}$ & Adhesion strength between active cells & $10.0$ \\ \hline
$\lambda_V$ & Volume constraint & $2.0$ \\ \hline
$V_t$ & Target volume & $100$ pixels \\ \hline
$\kappa$ & Motility force strength & $50.0$ \\ \hline 
$\tau$ & Persistence time & \textbf{2500} mcs \\ \hline
$\phi$ & Fraction of active cells & \textbf{0.25} \\ \hline
$\gamma$ & Alignment strength between active cells & \textbf{0.01} \\ \hline 
\end{tabular}
\end{table}

All cells in the simulation are identical, apart from their motility which is only applied to the motile cells. The adhesion energies $J_{\alpha_i, \alpha_j}$ are set to $10$ for all cell types. The clustering dynamics can therefore not be governed by cell sorting as in Ref.~\cite{Graner_CellSorting}. The target volume $V_t$ and the strength of the volume constraint $\lambda_V$ for the cells are set to $100$ pixels and $2.0$, respectively. The simulation box contains $400$ by $400$ pixels, so there are $1600$ cells included in the simulations. Periodic boundary conditions are applied.  
We assume that the fraction of motile cells $\phi$ is relatively small, with typically a value of $\phi=0.25$. We also ran simulations for a smaller fraction $\phi = 0.10$ and observed similar clustering behavior. These fractions are sufficiently large to allow motile cells to form clusters but not too large to let these motile clusters dominate the behavior of the primary tumor. The properties that govern the motility of the motile cells are the active force strength $\kappa$, the persistence time $\tau$, and the alignment strength $\gamma$. We set the active force strength $\kappa$ to $50$. This value is sufficiently high to induce cell migration but not too high as to only induce single cell migration \cite{Kabla_collectivecellmigration}. The persistence time and the alignment strength are varied such that we can studied their effect on cluster formation. The complete set of simulation parameters is provided in Table \ref{tab:simparams}.

To get an estimate for the typical values of the cells and their dynamics in the simulations, we use information about the dynamics of breast cancer cells in dense tissue \cite{Kim_UnjammingMigration_BreastCancerCells, West_DynamicsCancerousTissue}. A typical area of such breast cancer cells is $400$ $\si{\micro \meter}^2$ and therefore the size of a pixel in the simulation is $2~\si{\micro\meter}$ by $2~\si{\micro\meter}$. To determine the typical time scale, we can use the characteristic velocity of cells. Based on these velocities (see Supplementary Material for details), the time scale of $1$~mcs represents approximately $7$ seconds. 

We initialize the system with non-motile cells of $10$ by $10$ pixels and randomly change a fraction $\phi$ of those cells to be active. We start the simulation with the Hamiltonian in Eq.~\ref{eq:original_hamiltonian}, i.e., without the motility term, to allow the initial configuration to equilibrate. Since the cells are identical apart from the motile term, there is no clustering process happening during this initial phase of the simulation. After $2500$~mcs the cell shapes have equilibrated so the complete Hamiltonian in Eq.~\ref{eq:hamiltonian_term_active} is used and we start the analysis. The results are averaged over $200$ independent simulations and the data is stored every $200$~mcs. Since we do not know a priori where the motile cells are formed and whether they are formed randomly in space or concentrate in a small subdomain of the primary tumor, we also ran simulations with a different initial condition in which all motile cells start as a single strongly aligned cluster. While it is known that the spatial distribution of motile cells can influence clustering and migration behavior \cite{prasanna2024spatial}, we have verified that, in our simulations, the initial distribution of the motile cells does not affect the long-time steady state that we observe (see Supplementary Material). The steady state is reached after $60~000$~mcs for all parameters studied in this work. We therefore collect the relevant steady state data, indicated with a subscript $\infty$, for all simulations after this point in time.

\subsection{Characterizing clustering} \label{subsec:method_clustering}
We investigate four different scenarios to study the role of the dense environment and the cell-cell alignment interactions on the clustering of migrating cells:  
\begin{enumerate}
    \item Active cells in free space with alignment interactions ($\gamma > 0$);
    \item Active cells in a confluent layer with alignment interactions ($\gamma>0$);
    \item Active cells in free space in absence of alignment ($\gamma = 0$);
    \item Active cells in a confluent layer in absence of alignment ($\gamma = 0$).
\end{enumerate}
For each of these simulations, we characterize the collective motion of cells by measuring the global ordering of the active force and the clustering of motile cells. 

We measure the global ordering of the active force with an order parameter $P$ \cite{Vicsek_NovelTypePhaseTransition},
\begin{equation}
    P = \frac{1}{N} \left|\sum_{\sigma=1}^N \hat{p}_\sigma\right|,
    \label{eq:globalorderparameterP}
\end{equation}
where $N$ is the total number of motile cells and $\hat{p}_\sigma$ is the unit vector set by the cell's orientation angle $\theta_\sigma$ (Eq.~\ref{eq:angular_direction_both_pa_noise}). When the active force of all cells points in random directions the order parameter $P$ is equal to $0$, while its value is equal to $1$ when all polarity vectors point in the same direction. Note that the active force vector is different from the emerging velocity in the simulations, since the net movement of the cells is also affected by shape fluctuations and the presence of other cells. The order parameter serves to provide information about the underlying collective order. 

\begin{figure}
    \centering
    \includegraphics[width=0.85\linewidth]{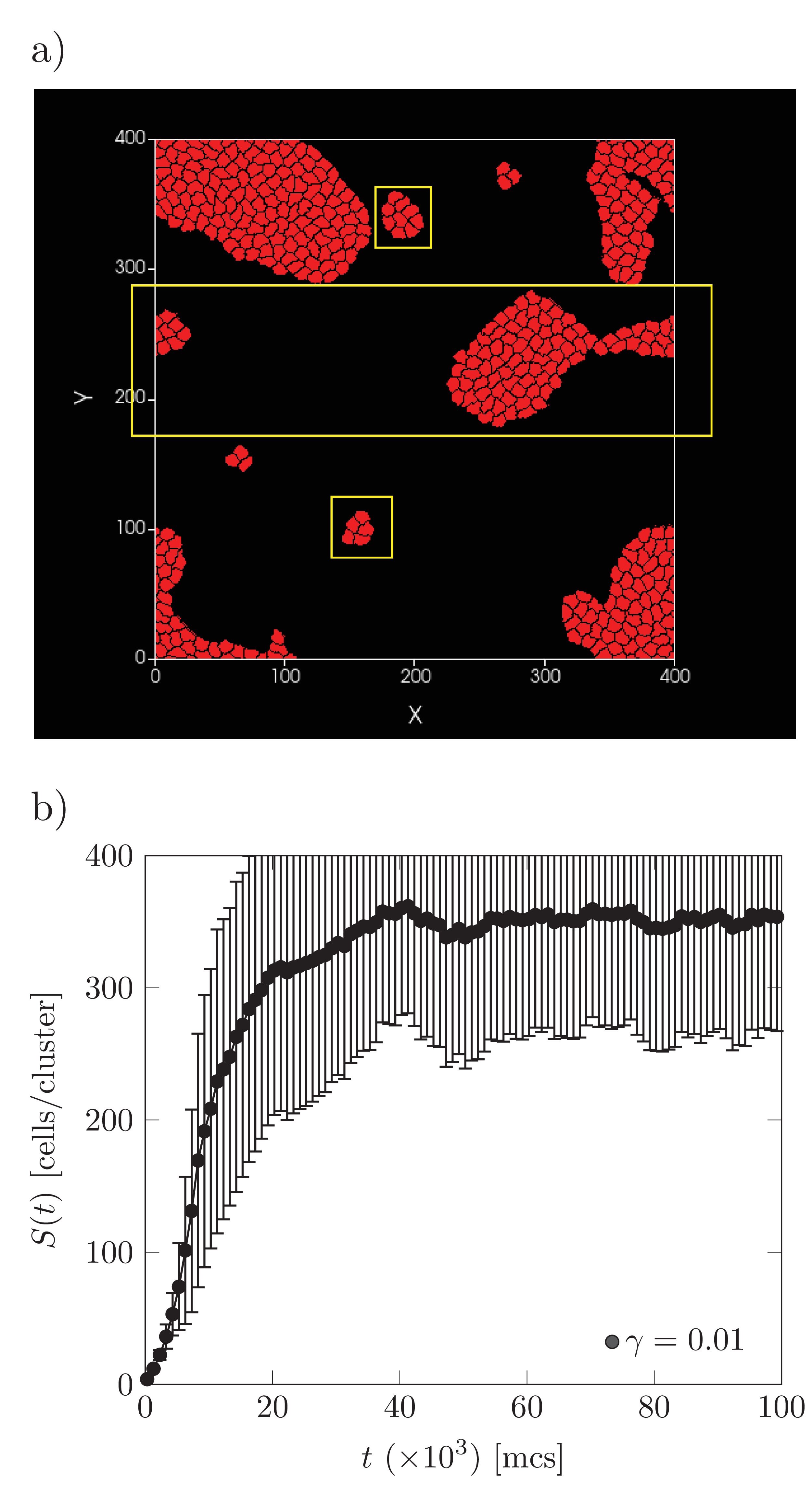}
    \caption{(a) Snapshot of the simulation with $400$ active cells for $\gamma = 0.01, \tau = 2500, \kappa = 50$ at $5000$~mcs. Three different clusters are encircled in yellow. (b) Mean cluster size as function of time, calculated with Eq.~\ref{eq:meanclustersize}. After an initial increase in the mean cluster size, the system evolves towards a steady state.}    
    \label{fig1:freespace_snapshot}
\end{figure}

Similarly, we measure the clusters in the simulations for the four distinct situations. Clusters are defined as groups of motile cells, as indicated by the yellow boxes in Fig.~\ref{fig1:freespace_snapshot}a, and the number of cells in a cluster is given by $S_i$. The mean cluster size $S$ is defined as 
\begin{equation}
    S = \frac{1}{N_c} \sum_{i=1}^{N_c} S_i = \frac{N}{N_c},
    \label{eq:meanclustersize}
\end{equation}
where $N$ is the number of active cells in the simulation, $N_c$ is the number of clusters, and $S_i$ is the number of cells in cluster $i$. Apart from the mean cluster size, we also study the steady-state cluster size distribution and the breaking and merging of clusters in time. The cluster size distribution is defined as $n_m m/N$, where $n_m$ is the number of clusters of size $m$. Essentially, it defines the probability of taking a cell and finding it in a cluster of size $m$. 

Finally, we determine the breaking and merging of cell clusters by comparing how the cells are distributed over the different clusters at different time steps. At time $t$, we check to which cluster each individual cell belongs to get the distribution of the motile cells over the clusters. At the consecutive time step $t'$, we perform the same analysis. We then compare how the composition of the clusters has changed between these consecutive time steps. For example, if a cluster consists of cells with cell number labels $\sigma$ equal to $4, 6$, and $12$ at time $t$, and these same cells are distributed over two clusters (a cluster with cells $4, 12$ and a cluster with cell $6$) at the later time $t'$ , we count a breaking event. Similarly, we count a merging event if cells that were not part of a cluster at time $t$ belong to the same cluster at time $t'$. The breaking and merging events are counted simultaneously, and inform us about the clustering dynamics of the motile cells in the simulation. 

\section{Results}
\subsection{Active cells in free space with alignment}
First, we consider the dynamics of motile cells in free space to benchmark the clustering behavior without the dense cellular environment. In this case the clustering is determined purely by the alignment interactions and the rotational diffusion. A snapshot of the simulation for $\gamma = 0.01$ and $\tau = 2500$~mcs at timestep $5000$~mcs is shown in Fig.~\ref{fig1:freespace_snapshot}a, where the system of cells is still evolving towards a steady state. The time evolution of the mean cluster size shows how the average number of cells per cluster grows as a function of time (Fig.~\ref{fig1:freespace_snapshot}b). The motile cells are initially randomly distributed in space, while most cells belong to the largest cluster in the steady state. 

\begin{figure*}
    \centering
    \includegraphics[width=\linewidth]{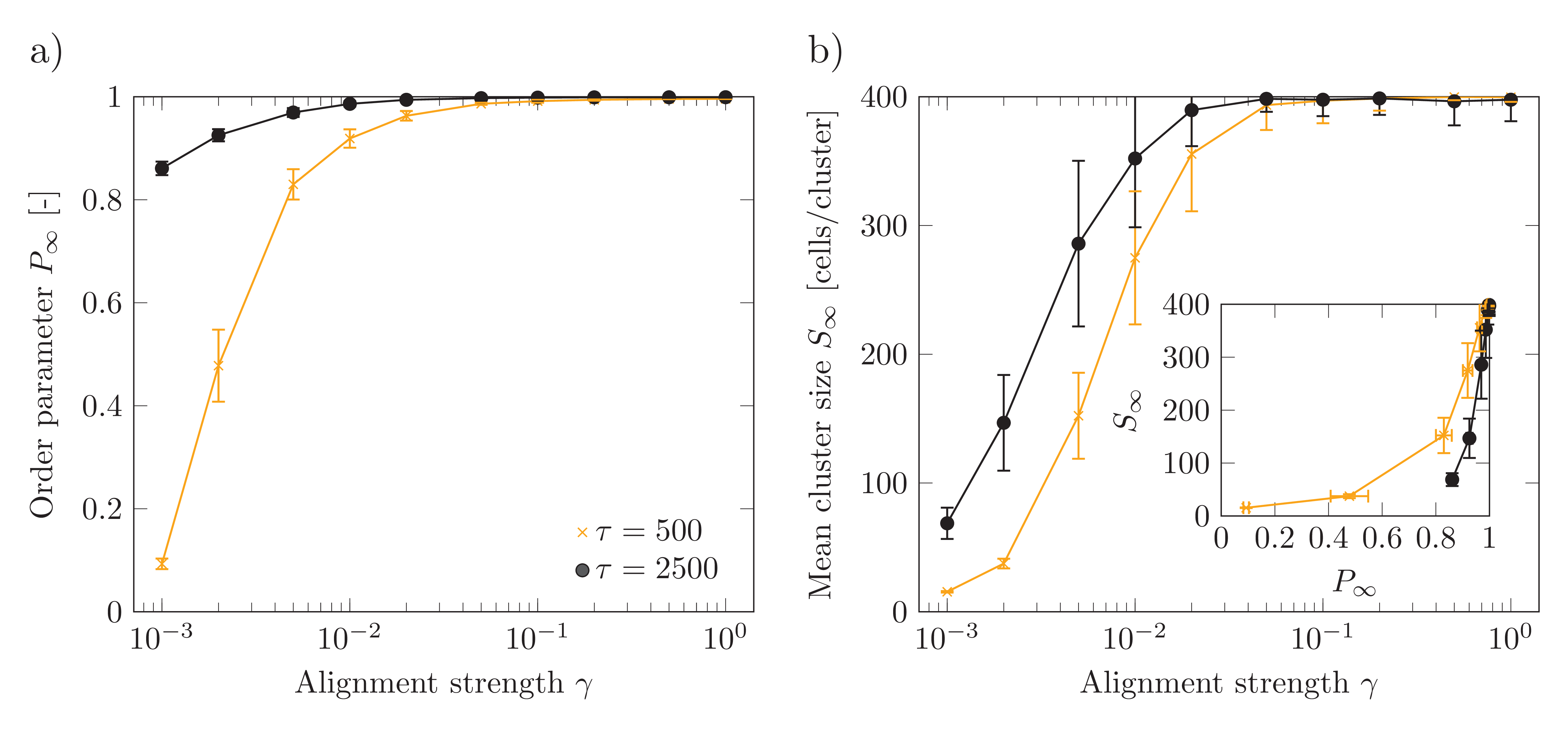}
\caption{(a) The global order parameter $P_{\infty}$ and (b) the mean cluster size $S_{\infty}$ as a function of the alignment strength $\gamma$ for $\tau = 500$~mcs (orange, square), $\tau = 2500$~mcs (black, circle) and $\tau = 4000$~mcs (blue, triangle). The simulations include $400$ motile cells that can migrate freely in space. (c) The correlation between $P_\infty$ and $S_\infty$. Decreasing the alignment strength decreases both the order parameter and the mean cluster size, and thereby leads to less collective behavior.}
\label{fig2:freespace_alignment}
\end{figure*}

We measure the global order parameter $P$ upon varying both the persistence time $\tau$ and the alignment strength $\gamma$ (see Fig.~\ref{fig2:freespace_alignment}a). As $\tau$ is increased, we observe a monotonic increase in the global ordering, fully consistent with the standard Vicsek model \cite{Vicsek_NovelTypePhaseTransition}. The order parameter also increases as we increase the alignment strength. For $\tau = 500$~mcs, we find a transition from a highly disordered state where $P$ is close to $0$ to an ordered state as $\gamma$ approaches $1$. Even though both parameters determine the underlying ordering of the active forces of the cells, we interpret the effect of the two parameters differently: the increase in alignment strength enhances the alignment of the active force orientations between neighboring cells, and thereby suppresses fluctuations in the active force between cells. This tends to increase the order parameter. Conversely, the persistence time acts as a rotational noise term and therefore enhances fluctuations in the active forces, leading to a decrease in the order parameter. 

Next, we turn to the mean cluster size in steady state. Fig.~\ref{fig2:freespace_alignment}b shows how the mean cluster size increases as we increase either $\gamma$ or $\tau$. When the order parameter is close to~$1$, one single cluster is formed. As we decrease the persistence time and the alignment strength, the mean cluster size decreases, indicating that there are more clusters present in the simulations. There is a also correlation between the mean cluster size and the global order parameter (see Fig.~\ref{fig2:freespace_alignment}c). These curves do not show a universal scaling behavior, however, implying that the global order parameter does not solely dictate the steady-state mean cluster size, and varying $\gamma$ and $\tau$ independently leads to different clustering behavior.

\subsection{Active cells in a confluent layer with alignment}
\begin{figure}
    \centering
    \includegraphics[width=0.85\linewidth]{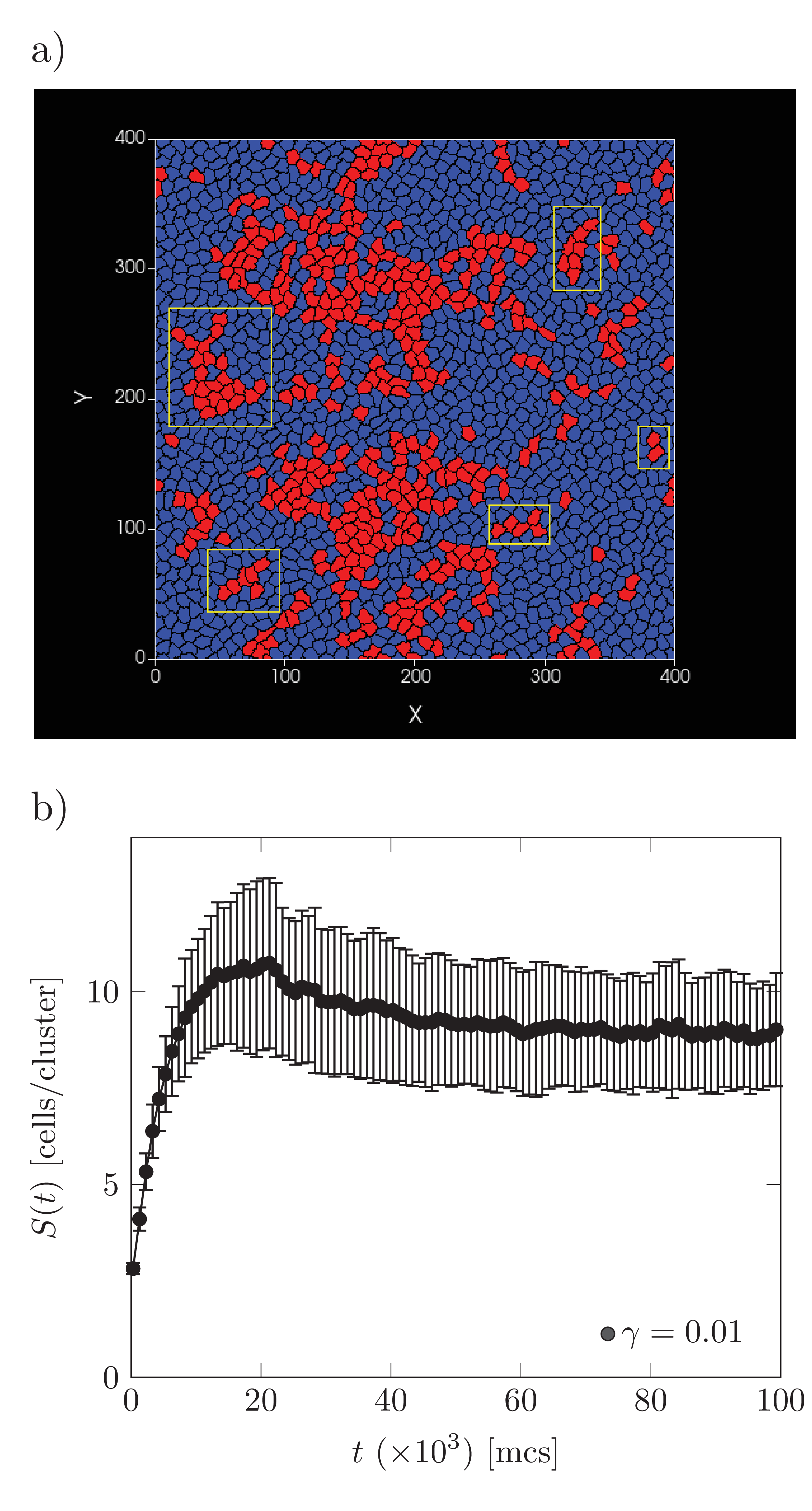}
    \caption{(a) Snapshot of the simulation for $\gamma = 0.01, \tau = 2500, \kappa = 50$ at $5000$~mcs. Five different clusters are encircles in yellow. (b) Mean cluster size $S(t)$ as a function of time, calculated with Eq.~\ref{eq:meanclustersize}. After an initial increase, the mean cluster size evolves towards a steady state. The mean cluster size shows an optimum due to the presence of the non-motile cells.}
\label{fig3:confluent_snapshot}
\end{figure}

We now consider the more realistic case of a heterogeneous confluent layer with both motile and non-motile cells. Fig.~\ref{fig3:confluent_snapshot}a shows a snapshot of the simulation for $\gamma~=~0.01$ and $\tau = 2500$~mcs at timestep $5000$~mcs. The clusters that are formed after $5000$~mcs are different from those in free space with the same set of parameters. The clusters have a different morphology and there are more smaller clusters present. Video S1 shows the dynamic evolution of clusters for $t=0$~mcs to $t=22~ 500$~mcs. The clusters are more strand-like and irregularly shaped, similar to active cell structures observed in Refs.~\cite{Kabla_collectivecellmigration, Hallou_tumorheterogeneitypromotescollectiveinvasion, Sandersius_EmergentCellAndTissueDynamics}. The cluster morphology is shaped by the cells coordinating their motion with their neighbors but the cell's migration direction is also restricted by the non-motile cells in the surroundings. Indeed, two strongly aligned clusters in the two different surroundings studied in this paper evolve very differently: while the motile cells in free space migrate collectively (video S2), the cluster in the confluent layer scatters and forms motile-cell strands (video S3). This phenomenon leads to a significantly smaller mean cluster size than in free space over the same time span, and the magnitude of the mean cluster size does not grow as fast (see Fig.~\ref{fig3:confluent_snapshot}b). Moreover, there is an overshoot in the mean cluster size before eventually reaching a steady state. The occurrence of such overshoots has been demonstrated as a characteristic of the non-equilibrium nature of biological systems \cite{Jia_overshoot_in_biologicalsystems}. Even though the cell dynamics of motile cells is governed by the same set of parameters as in free space, their dynamic behavior changes when migrating in a dense confluent layer. 

\begin{figure*}
    \centering
    \includegraphics[width=\textwidth]{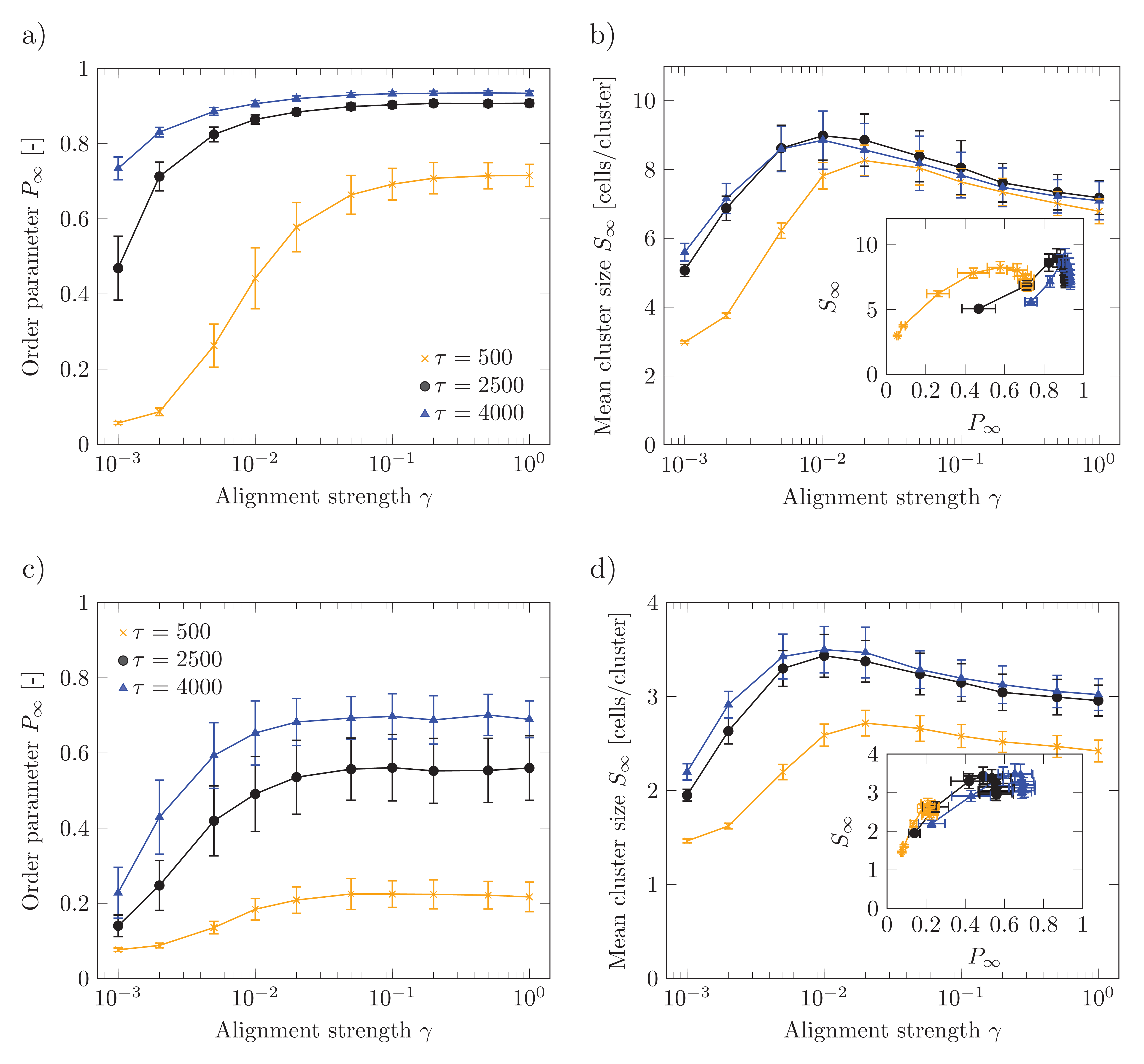}
    \caption{(a) The global order parameter $P_{\infty}$ and (b) the mean cluster size $S_{\infty}$ at a fraction of $0.25$ as a function of the alignment strength $\gamma$ for persistence times $\tau = 500$~mcs (orange, square), $\tau = 2500$~mcs (black, circle) and $\tau = 4000$~mcs (blue, triangle). (c) The correlation between $P_\infty$ and $S_\infty$. The order parameter shows the same trends upon changing $\gamma, \tau$ as for motile cells under free space conditions. For the mean cluster size, we see an optimum for intermediate alignment strength. (d, e, f) The same data for an active fraction $\phi = 0.10$. }
    \label{fig4:confluent_PS}
\end{figure*}

Returning to the steady-state analysis, we observe that the order parameter $P$ shows the same trends upon changing $\gamma$ and $\tau$ as for the simulations in free space (see Fig.~\ref{fig4:confluent_PS}a). The absolute value of the order parameters is slightly lower, but decreasing either the persistence time or the alignment strength decreases the overall global ordering in the system as was the case for active cells in free space. The small decline in $P$ can be explained by the fact that clusters are smaller and therefore fewer interactions between cells occur. Indeed, when we consider the confluent layer with a smaller number of active cells ($\phi=0.10$), and as a consequence, fewer interactions between active cells, the order parameter also becomes smaller (Fig.~\ref{fig4:confluent_PS}d). Thus, the non-motile cells have a minimal effect on the global ordering of the active forces. 

While the order parameter hardly changes in the confluent layer in steady-state, the mean cluster size shows that the clustering of motile cells in a confluent layer is significantly different from that in free space. Fig.~\ref{fig4:confluent_PS}b shows that the mean cluster size scales non-monotonically with the alignment strength for different values of $\tau$. For small alignment strengths ($\gamma < 10^{-2}$), there is a regime where both the mean cluster size $S$ and the order parameter $P$ are relatively small. On the other side of the spectrum where the alignment is strongest ($\gamma$ close to $1$), we note that the mean cluster size is smaller compared to the intermediate regime ($\gamma \approx 10^{-2}$) as well. Fig.~\ref{fig4:confluent_PS}c shows that the correlation between the order parameter and the mean cluster size no longer scales monotonically. We have verified that the steady-state values do not depend on the initial condition, see Supplementary Material. Fig.~\ref{fig4:confluent_PS}e shows that the same trend is also observed when $\phi=0.10$. The presence of the non-motile cells in the confluent layer strongly affects the mean cluster size and leads to the most significant decrease in the mean cluster size for large alignment strength $\gamma$. 

Based on the trends observed for $P$ and $S$, our results show that the clustering of motile cells in the confluent layer is different from that in free space even though the global ordering of the active force remains almost unaltered. A similar phenomenon of different clustering phases with similar ordering has been detected in the original Vicsek model recently \cite{Miyahara_VicsekMeetsDBSCAN}. In our confluent layer, three aspects (instead of two in free space) determine the emerging clustering dynamics. Apart from the persistence time $\tau$ and the alignment strength $\gamma$, we also need to consider the non-motile cells comprising the dense cellular environment. As discussed before, the persistence time and the alignment strength have distinct effects on the clustering behavior: Based on the results in free space, for which the mean cluster size increases with $\gamma$, we can interpret enhanced cell-cell alignment as a promoter of collective cell migration, suppressing the breaking of clusters once formed. Decreasing the persistence time (increasing rotational noise) leads to a decrease in the mean cluster size, thus seemingly counteracting alignment effects and promoting the breaking of clusters. 
These two effects are now in competition with the presence of the non-motile cells, which manifestly decrease the mean cluster size. Such non-motile cells could potentially introduce an additional noise term by manifesting themselves as obstacles when cells are clustering. 

We can discern the interplay between the three effects by investigating the results for $\phi=0.10$ and $0.25$. For $\phi = 0.25$, the number of active cells is relatively high and frequent interactions between cells increase the alignment. Larger persistence times lead to larger clusters in the low $\gamma$-regime, similar to free space. In the high $\gamma$-regime, on the other hand, the persistence time hardly affects the mean cluster size. The alignment and the presence of the cellular environment must dominate the clustering behavior in this regime. The rotational noise is strongly suppressed while the non-motile cells introduce an additional noise term that affects the emerging clustering dynamics. At a lower fraction of active cells, $\phi = 0.10$, the number of interactions between active cells is lowered and thereby the overall alignment effect decreases. Since the persistence time is a single-cell property, it is not affected by the fraction of active cells in the confluent layer. Therefore, cells have fewer interactions when $\phi=0.10$ so the alignment is less dominant and the rotational noise has a more noticable effect. We can observe this effect by the increase in the mean cluster size upon increasing $\tau$ over the entire range of $\gamma$. For $\phi=0.10$, the environment again plays the most prominent role for large alignment strength by decreasing the mean cluster size for large $\gamma$. Overall, we observe a competition between the alignment strength and the persistence time, and we find that the dense cellular environment strongly suppresses the formation of clusters where the most notable effect is visible for large alignment strengths. 

\begin{figure}
    \centering
    \includegraphics[width=\linewidth]{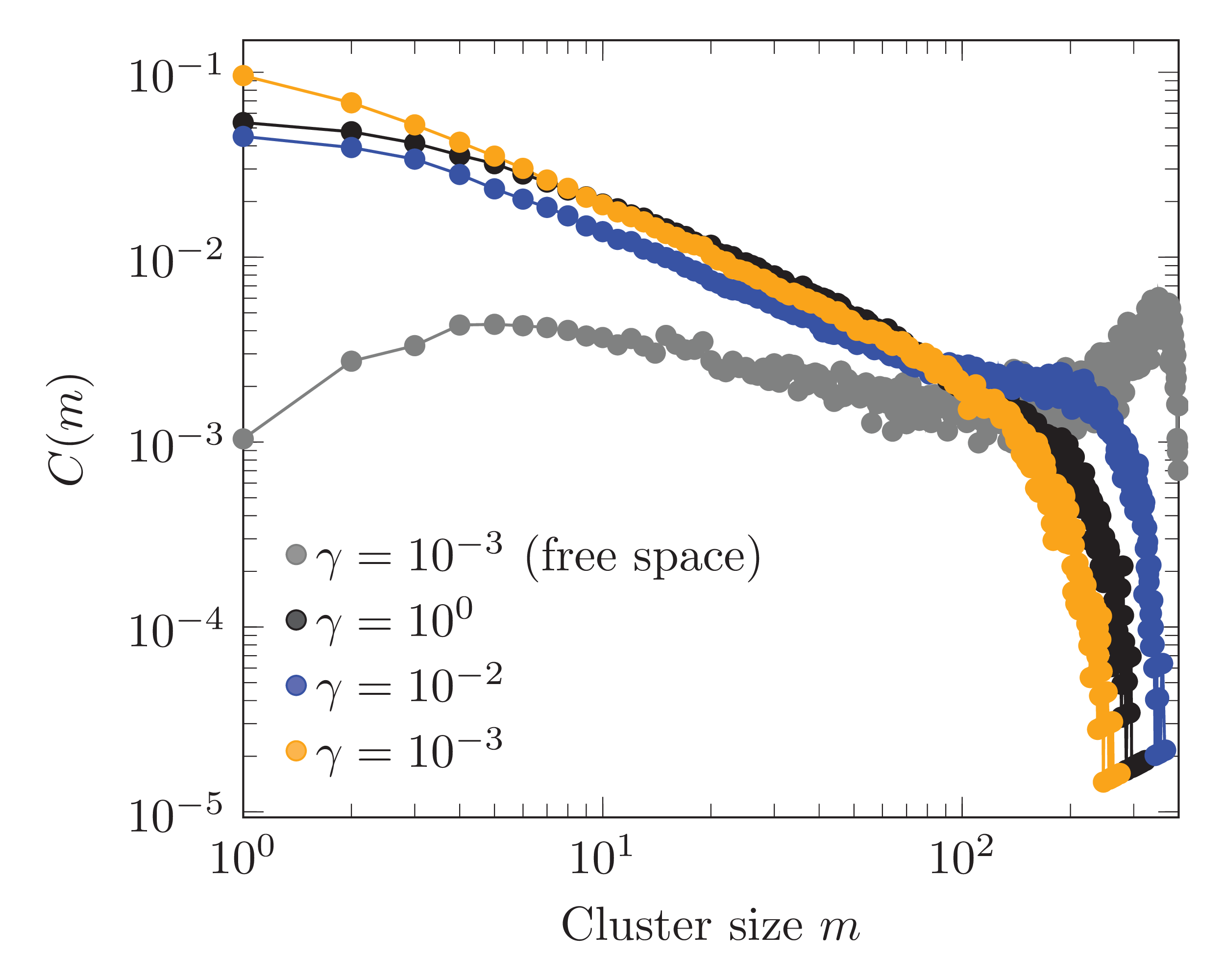}
    \caption{Cluster size distribution $C(m)$ for $\tau = 2500$~mcs and three different values for $\gamma = 1, 10^{-2}, 10^{-3}$ in the confluent layer. The distribution in free space for $\gamma=10^{-3}$ is included in gray. The presence of smaller clusters is apparent in the confluent layer. Moreover, the largest clusters form at intermediate alignment strength. }\label{fig5:confluent_CSD}
\end{figure}

To get more information about the clusters that are formed in the different regimes of $\gamma$, we characterize the cluster size distribution in Fig.~\ref{fig5:confluent_CSD}. First of all, the cluster size distribution is significantly different for motile cells in a confluent layer compared to the case in free space. In the dense cellular environment, single cells are detected more frequently. Interestingly, around the optimum in the mean cluster size ($\gamma = 10^{-2}$), the cluster size distribution of Fig.~\ref{fig5:confluent_CSD} reveals that there is a relative abundance of large clusters (relative with respect to other values of $\gamma$) and a relatively low number of smaller clusters. At this value of $\gamma$, we also see that the probability of forming large clusters initially decreases but peaks for the large clusters. This trend, in which many small clusters appear in tandem with large cell clusters, has also been found in a theoretical model by Peruani and co-workers \cite{Peruani_kineticmodelandscalingproperties,peruani_clusterdynamics_2010, Starrus_clusterdynamics_with_omega}. In their model, the breaking of clusters is fully determined by single-particle detachment. Although our simulations also contain multicellular cluster detachment, the similarities between our simulations and their theoretical predictions indeed show that single-cell and small-cluster detachments are an important breaking mechanism for cell clusters in the confluent layer. 
 
\subsection{Clustering without alignment}
The results presented so far have shown that both the alignment strength and persistence time, as well as the dense cellular environment, determine how clusters form in the primary tumor. To elucidate the role of the dense environment in absence of these alignment interactions, we set $\gamma = 0$ and focus on the clustering dynamics solely due to the environment.  

\begin{figure*}
    \centering
    \includegraphics[width=\linewidth]{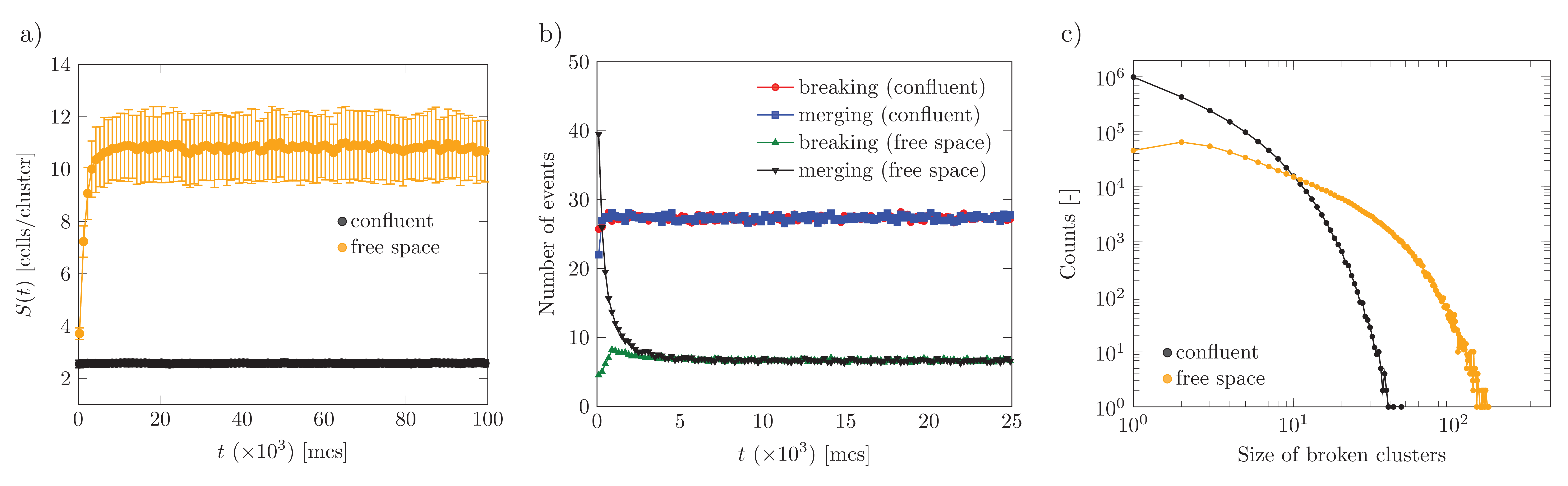}
\caption{(a) Mean cluster size in time, (b) the number of breaking and merging events in time and (c) the cluster size distribution after breaking in the steady state for motile cells in both free space and the confluent layer for $\tau = 2500$~mcs and $\gamma = 0.0$, i.e. without any alignment interactions. The confluent layer enhances the breaking and merging of clusters, promoting the formation of smaller clusters and thereby decreasing the mean cluster size. }
\label{fig6:noalignment}
\end{figure*}

Figure \ref{fig6:noalignment}a compares the mean cluster size as a function of time for confluent and free-space conditions, starting from a random distribution of active cells with $\gamma = 0$. It can be seen that the mean cluster size in the confluent layer remains very small and does not change in time, while in free-space conditions the motile cells spontaneously undergo clustering at short times, resulting in a much larger mean cluster size. This initial clustering in free space can be attributed to cell-cell adhesion effects, whereby two cells get close together to decrease the total contact surface with the medium. In the confluent layer, however, the mean cluster size remains identical to the random initial configuration, implying that motile cells remain randomly distributed in space when cell-cell alignment interactions are absent. In this case, small clusters are only formed due to random transient encounters. From comparing the case $\gamma=0$ with $\gamma > 0$, we can thus conclude that alignment is necessary to form motile cell clusters in a dense cellular environment. 

To understand how the dynamics of the clusters evolves in time (in the absence of cell-cell alignment), we plot the number of breaking and merging events in Fig.~\ref{fig6:noalignment}b as defined in Sec.~\ref{subsec:method_clustering}. The breaking and merging of clusters is measured by the changes in the distribution of the motile cells over the different clusters between consecutive time steps. The number of breaking and merging events increases in the dense cellular environment compared to free space. Even though the mean cluster size in the confluent layer does not change in steady state, the clusters are highly dynamic. Clusters are constantly changing their composition and dynamically forming and breaking in time. Fig.~\ref{fig6:noalignment}c shows the size distribution of clusters after breaking in steady state. In the confluent layer, small clusters are detected more frequently. Thus, in absence of any alignment, the confluent layer enhances the breaking and merging of clusters, promoting the formation of smaller clusters and thereby decreasing the mean cluster size as compared to the same motile cells in free space.

Finally, we can compare the results for $\gamma = 0$ with the results for $\gamma > 0$ to hypothesize how the interplay between the cell-cell alignment interactions and the dense cellular environment affects the breaking and merging of clusters during the clustering process. Our simulations have shown that the motile clusters in the confluent layer fragment more significantly than in free space when cells interact with the same alignment strength in both surroundings. This breaking of clusters in the confluent layer is inherently different from breaking of clusters in free space where clusters fragment due to a lack of alignment. The non-motile cells in the dynamic cellular environment introduce an additional type of noise in the system. The fragmentation of clusters shows similarities with the disruption of local ordering due to the presence of obstacles. These obstacles can either manifest themselves as a small number of non-aligning agents \cite{Yllanes_howmanydissenters} or as physical (non-motile) obstacles \cite{Chepizho_activeparticles_inheterogenous_media, Martinez_CollectiveBehaviorwithandwithouObstacles}. The non-motile cells in the confluent layer effectively acts as a destructive obstacle that introduces a different cluster breaking mechanism compared to the lack of coordinated motion. Together with the dynamics induced by introducing cell-cell alignment interactions, the clustering of motile cells in a dense tumor environment is a complex process by which multiple breaking mechanisms, affected by multiple sources of noise, simultaneously play a role.

\section{Conclusion}
In this work, we have investigated the role of cell-cell alignment in the formation of migratory cell clusters in the bulk of a model primary tumor. We simulated the cell dynamics of a small fraction of motile cells in a dense non-motile cellular environment using the CPM. To identify the role of the cell-cell alignment interactions in such a confluent environment, we studied how the coordinated motion of cells, or the lack thereof, determines how actively-driven cells in dense cellular environments form clusters. We focus on the pre-detachment phase of cell migration, but we expect that these motile clusters could develop inside dense tumors and initiate the formation of metastasis-seeding CTC clusters in later stages.

Our computational model has allowed us to study both the effect of cell-cell alignment interactions, and the role of the dense cellular environment, separately. In absence of any alignment interactions, the cellular environment promotes the break-up of small clusters. The non-motile cells act as an impediment to the clustering process. When cell-cell alignment interactions are included, next to the cell-intrinsic persistence time $\tau$, cell dynamics is also governed by a cell-cell alignment interaction strength $\gamma$. Both affect the net ordering of the coordinated motion, but the parameters $\gamma$ and $\tau$ affect the clustering that is observed in the confluent layer differently: while increasing the alignment strength promotes alignment and thereby the formation of clusters, increasing the persistence time suppresses the formation of larger clusters by hindering effective alignment. One of the key findings of our work is an optimum in the mean cluster size upon changing $\gamma$ in the confluent layer. At the optimum, with an alignment strength around $\gamma = 10^{-2}$, the largest cell clusters are formed and smaller clusters are detected less frequently compared to other $\gamma$ values. The emergence of the optimum is independent of the initial spatial distribution of the motile cells in the confluent layer. The change in the clustering behavior upon increasing $\gamma$ is due to a transition from a regime where breaking is dominated by a lack of alignment interactions to a regime where breaking is dominated by the scattering of motile cells on non-motile obstacles. The formation of large clusters is strongly suppressed in the confluent layer for large $\gamma$. These results suggest that the formation of actively migrating cells clusters inside the primary tumor is impeded by many distinct effects at play simultaneously. This observation could be an important factor to rationalize the fact that the CTC clusters that are detected in experiments are relatively small. 

Reviewing prior work on similar systems, we note that the emergence of an optimum has been observed before in cellular systems in e.g.\ Ref.~\cite{roy2021intermediate}. The origin of that optimum, however, is different from that in our work as it results from direct cell-cell interactions, rather than from the alignment interactions we consider. This prior work does show that cell-cell adhesion is also a critical factor in the formation and migration of cell clusters, and is therefore a relevant parameter to include in future research. Another potentially impactful factor in the clustering process is cell division. Our results are valid in the regime where cell division is either absent, or slow compared to the time necessary to reach a steady state; in this case the spatial evolution could be considered in a quasi-steady state at all times. More work is needed to explore clustering dynamics when the time-to-steady-state is shorter than, or comparable to, the division time. Finally, our results suggest that the initial spatial distribution of the motile cells does not affect the steady-state mean cluster size. Previous computational work has shown that this spatial distribution does become important when cells detach from the primary tumor \cite{prasanna2024spatial}. This discrepancy highlights the existence of multiple clustering mechanisms, and should inform future research exploring how alignment interactions affect cell dynamics also in the later stages of the metastatic cascade.

Finally, we note that our current work considers a minimally heterogeneous 2D tumor consisting exclusively of motile and non-motile cells. These two cell types represent a mesenchymal and epithelial cell phenotype, respectively. One important aspect that has only been taken into account partially with this set-up is the fact that the epithelial-to-mesenchymal transition is usually a hybrid transition, and primary tumors usually contain cells on a continuous spectrum between epithelial and mesenchymal, rather than two distinct phenotypes. Such cells express both epithelial and mesenchymal markers and therefore can have both epithelial and mesenchymal properties \cite{Jolly_ImplicationsOfHybridEMT}. When cells detach from the primary tumor, these hybrid states play a critical role in the formation of cell clusters. Computer simulations have shown that the metastasizing clusters also contains multiple cell phenotypes similar to the primary tumor core \cite{Mukherjee_ClusterSizeDistribution, Bocci_BiophysicalModel_uncovers_CSD_acrossCancer}. These papers also emphasize that the sizes of the clusters during metastasis are relatively small, on the order of a few cells. Our work suggests a similar size distribution inside the tumor core. We speculate that the motile cell clusters in our work may pull along the less motile cells in later stages of cancer progression, and thereby initialize the formation of heterogeneous cell clusters that eventually detach from the primary tumor core. Further research is necessary to reveal how the clusters that form inside a primary tumor will initiate the formation of circulating tumor cell clusters.  

\begin{acknowledgments}
We thank Vincent E.~Debets for his critical reviewing of the manuscript. This work has been financially supported by the Dutch Research Council (NWO) through the ENW-XL project ``Active Matter Physics of Collective Metastasis" (OCENW.GROOT.2019.022).
\end{acknowledgments}

\appendix

\bibliography{paper_review}

\begin{thebibliography}{60}%
\makeatletter
\providecommand \@ifxundefined [1]{%
 \@ifx{#1\undefined}
}%
\providecommand \@ifnum [1]{%
 \ifnum #1\expandafter \@firstoftwo
 \else \expandafter \@secondoftwo
 \fi
}%
\providecommand \@ifx [1]{%
 \ifx #1\expandafter \@firstoftwo
 \else \expandafter \@secondoftwo
 \fi
}%
\providecommand \natexlab [1]{#1}%
\providecommand \enquote  [1]{``#1''}%
\providecommand \bibnamefont  [1]{#1}%
\providecommand \bibfnamefont [1]{#1}%
\providecommand \citenamefont [1]{#1}%
\providecommand \href@noop [0]{\@secondoftwo}%
\providecommand \href [0]{\begingroup \@sanitize@url \@href}%
\providecommand \@href[1]{\@@startlink{#1}\@@href}%
\providecommand \@@href[1]{\endgroup#1\@@endlink}%
\providecommand \@sanitize@url [0]{\catcode `\\12\catcode `\$12\catcode `\&12\catcode `\#12\catcode `\^12\catcode `\_12\catcode `\%12\relax}%
\providecommand \@@startlink[1]{}%
\providecommand \@@endlink[0]{}%
\providecommand \url  [0]{\begingroup\@sanitize@url \@url }%
\providecommand \@url [1]{\endgroup\@href {#1}{\urlprefix }}%
\providecommand \urlprefix  [0]{URL }%
\providecommand \Eprint [0]{\href }%
\providecommand \doibase [0]{https://doi.org/}%
\providecommand \selectlanguage [0]{\@gobble}%
\providecommand \bibinfo  [0]{\@secondoftwo}%
\providecommand \bibfield  [0]{\@secondoftwo}%
\providecommand \translation [1]{[#1]}%
\providecommand \BibitemOpen [0]{}%
\providecommand \bibitemStop [0]{}%
\providecommand \bibitemNoStop [0]{.\EOS\space}%
\providecommand \EOS [0]{\spacefactor3000\relax}%
\providecommand \BibitemShut  [1]{\csname bibitem#1\endcsname}%
\let\auto@bib@innerbib\@empty
\bibitem [{\citenamefont {Dillekås}\ \emph {et~al.}(2019)\citenamefont {Dillekås}, \citenamefont {Rogers},\ and\ \citenamefont {Straume}}]{Dillekas_deathsduetometastasis}%
  \BibitemOpen
  \bibfield  {author} {\bibinfo {author} {\bibfnamefont {H.}~\bibnamefont {Dillekås}}, \bibinfo {author} {\bibfnamefont {M.~S.}\ \bibnamefont {Rogers}},\ and\ \bibinfo {author} {\bibfnamefont {O.}~\bibnamefont {Straume}},\ }\bibfield  {title} {\bibinfo {title} {Are 90\% of deaths from cancer caused by metastases?},\ }\href {https://doi.org/10.1002/cam4.2474} {\bibfield  {journal} {\bibinfo  {journal} {Cancer Medicine}\ }\textbf {\bibinfo {volume} {8}},\ \bibinfo {pages} {5574} (\bibinfo {year} {2019})}\BibitemShut {NoStop}%
\bibitem [{\citenamefont {Friedl}\ and\ \citenamefont {Wolf}(2009)}]{Friedl_plasticitycellmigration}%
  \BibitemOpen
  \bibfield  {author} {\bibinfo {author} {\bibfnamefont {P.}~\bibnamefont {Friedl}}\ and\ \bibinfo {author} {\bibfnamefont {K.}~\bibnamefont {Wolf}},\ }\bibfield  {title} {\bibinfo {title} {{Plasticity of cell migration: a multiscale tuning model}},\ }\href {https://doi.org/10.1083/jcb.200909003} {\bibfield  {journal} {\bibinfo  {journal} {Journal of Cell Biology}\ }\textbf {\bibinfo {volume} {188}},\ \bibinfo {pages} {11} (\bibinfo {year} {2009})}\BibitemShut {NoStop}%
\bibitem [{\citenamefont {Kang}\ \emph {et~al.}(2021)\citenamefont {Kang}, \citenamefont {Ferruzzi}, \citenamefont {Spatarelu}, \citenamefont {Han}, \citenamefont {Sharma}, \citenamefont {Koehler}, \citenamefont {Mitchel}, \citenamefont {Khan}, \citenamefont {Butler}, \citenamefont {Roblyer}, \citenamefont {Zaman}, \citenamefont {Park}, \citenamefont {Guo}, \citenamefont {Chen}, \citenamefont {Pegoraro},\ and\ \citenamefont {Fredberg}}]{Kang_novelPhaseDiagram_tumorInvasion}%
  \BibitemOpen
  \bibfield  {author} {\bibinfo {author} {\bibfnamefont {W.}~\bibnamefont {Kang}}, \bibinfo {author} {\bibfnamefont {J.}~\bibnamefont {Ferruzzi}}, \bibinfo {author} {\bibfnamefont {C.-P.}\ \bibnamefont {Spatarelu}}, \bibinfo {author} {\bibfnamefont {Y.~L.}\ \bibnamefont {Han}}, \bibinfo {author} {\bibfnamefont {Y.}~\bibnamefont {Sharma}}, \bibinfo {author} {\bibfnamefont {S.~A.}\ \bibnamefont {Koehler}}, \bibinfo {author} {\bibfnamefont {J.~A.}\ \bibnamefont {Mitchel}}, \bibinfo {author} {\bibfnamefont {A.}~\bibnamefont {Khan}}, \bibinfo {author} {\bibfnamefont {J.~P.}\ \bibnamefont {Butler}}, \bibinfo {author} {\bibfnamefont {D.}~\bibnamefont {Roblyer}}, \bibinfo {author} {\bibfnamefont {M.~H.}\ \bibnamefont {Zaman}}, \bibinfo {author} {\bibfnamefont {J.-A.}\ \bibnamefont {Park}}, \bibinfo {author} {\bibfnamefont {M.}~\bibnamefont {Guo}}, \bibinfo {author} {\bibfnamefont {Z.}~\bibnamefont {Chen}}, \bibinfo {author} {\bibfnamefont {A.~F.}\ \bibnamefont {Pegoraro}},\ and\ \bibinfo {author} {\bibfnamefont
  {J.~J.}\ \bibnamefont {Fredberg}},\ }\bibfield  {title} {\bibinfo {title} {A novel jamming phase diagram links tumor invasion to non-equilibrium phase separation},\ }\href {https://doi.org/10.1016/j.isci.2021.103252} {\bibfield  {journal} {\bibinfo  {journal} {iScience}\ }\textbf {\bibinfo {volume} {24}},\ \bibinfo {pages} {103252} (\bibinfo {year} {2021})}\BibitemShut {NoStop}%
\bibitem [{\citenamefont {Giuliano}\ \emph {et~al.}(2018)\citenamefont {Giuliano}, \citenamefont {Shaikh}, \citenamefont {Lo}, \citenamefont {Arpino}, \citenamefont {De~Placido}, \citenamefont {Zhang}, \citenamefont {Cristofanilli}, \citenamefont {Schiff},\ and\ \citenamefont {Trivedi}}]{Giuliano_CTC_VillagetoMetastasize}%
  \BibitemOpen
  \bibfield  {author} {\bibinfo {author} {\bibfnamefont {M.}~\bibnamefont {Giuliano}}, \bibinfo {author} {\bibfnamefont {A.}~\bibnamefont {Shaikh}}, \bibinfo {author} {\bibfnamefont {H.~C.}\ \bibnamefont {Lo}}, \bibinfo {author} {\bibfnamefont {G.}~\bibnamefont {Arpino}}, \bibinfo {author} {\bibfnamefont {S.}~\bibnamefont {De~Placido}}, \bibinfo {author} {\bibfnamefont {X.~H.}\ \bibnamefont {Zhang}}, \bibinfo {author} {\bibfnamefont {M.}~\bibnamefont {Cristofanilli}}, \bibinfo {author} {\bibfnamefont {R.}~\bibnamefont {Schiff}},\ and\ \bibinfo {author} {\bibfnamefont {M.~V.}\ \bibnamefont {Trivedi}},\ }\bibfield  {title} {\bibinfo {title} {{Perspective on Circulating Tumor Cell Clusters: Why It Takes a Village to Metastasize}},\ }\href {https://doi.org/10.1158/0008-5472.CAN-17-2748} {\bibfield  {journal} {\bibinfo  {journal} {Cancer Research}\ }\textbf {\bibinfo {volume} {78}},\ \bibinfo {pages} {845} (\bibinfo {year} {2018})}\BibitemShut {NoStop}%
\bibitem [{\citenamefont {Chemi}\ \emph {et~al.}(2021)\citenamefont {Chemi}, \citenamefont {Mohan}, \citenamefont {Guevara}, \citenamefont {Clipson}, \citenamefont {Rothwell},\ and\ \citenamefont {Dive}}]{Chemi_EarlyDissemination_of_CTCs}%
  \BibitemOpen
  \bibfield  {author} {\bibinfo {author} {\bibfnamefont {F.}~\bibnamefont {Chemi}}, \bibinfo {author} {\bibfnamefont {S.}~\bibnamefont {Mohan}}, \bibinfo {author} {\bibfnamefont {T.}~\bibnamefont {Guevara}}, \bibinfo {author} {\bibfnamefont {A.}~\bibnamefont {Clipson}}, \bibinfo {author} {\bibfnamefont {D.~G.}\ \bibnamefont {Rothwell}},\ and\ \bibinfo {author} {\bibfnamefont {C.}~\bibnamefont {Dive}},\ }\bibfield  {title} {\bibinfo {title} {Early dissemination of circulating tumor cells: Biological and clinical insights},\ }\href {https://doi.org/10.3389/fonc.2021.672195} {\bibfield  {journal} {\bibinfo  {journal} {Frontiers in Oncology}\ }\textbf {\bibinfo {volume} {11}},\ \bibinfo {pages} {672195} (\bibinfo {year} {2021})}\BibitemShut {NoStop}%
\bibitem [{\citenamefont {Aceto}\ \emph {et~al.}(2014)\citenamefont {Aceto}, \citenamefont {Bardia}, \citenamefont {Miyamoto}, \citenamefont {Donaldson}, \citenamefont {Wittner}, \citenamefont {Spencer}, \citenamefont {Yu}, \citenamefont {Pely}, \citenamefont {Engstrom}, \citenamefont {Zhu}, \citenamefont {Brannigan}, \citenamefont {Kapur}, \citenamefont {Stott}, \citenamefont {Shioda}, \citenamefont {Ramaswamy}, \citenamefont {Ting}, \citenamefont {Lin}, \citenamefont {Toner}, \citenamefont {Haber},\ and\ \citenamefont {Maheswaran}}]{Aceta_CTCsareprecursorofmetastasis}%
  \BibitemOpen
  \bibfield  {author} {\bibinfo {author} {\bibfnamefont {N.}~\bibnamefont {Aceto}}, \bibinfo {author} {\bibfnamefont {A.}~\bibnamefont {Bardia}}, \bibinfo {author} {\bibfnamefont {D.~T.}\ \bibnamefont {Miyamoto}}, \bibinfo {author} {\bibfnamefont {M.~C.}\ \bibnamefont {Donaldson}}, \bibinfo {author} {\bibfnamefont {B.~S.}\ \bibnamefont {Wittner}}, \bibinfo {author} {\bibfnamefont {J.~A.}\ \bibnamefont {Spencer}}, \bibinfo {author} {\bibfnamefont {M.}~\bibnamefont {Yu}}, \bibinfo {author} {\bibfnamefont {A.}~\bibnamefont {Pely}}, \bibinfo {author} {\bibfnamefont {A.}~\bibnamefont {Engstrom}}, \bibinfo {author} {\bibfnamefont {H.}~\bibnamefont {Zhu}}, \bibinfo {author} {\bibfnamefont {B.~W.}\ \bibnamefont {Brannigan}}, \bibinfo {author} {\bibfnamefont {R.}~\bibnamefont {Kapur}}, \bibinfo {author} {\bibfnamefont {S.~L.}\ \bibnamefont {Stott}}, \bibinfo {author} {\bibfnamefont {T.}~\bibnamefont {Shioda}}, \bibinfo {author} {\bibfnamefont {S.}~\bibnamefont {Ramaswamy}}, \bibinfo {author} {\bibfnamefont {D.~T.}\
  \bibnamefont {Ting}}, \bibinfo {author} {\bibfnamefont {C.~P.}\ \bibnamefont {Lin}}, \bibinfo {author} {\bibfnamefont {M.}~\bibnamefont {Toner}}, \bibinfo {author} {\bibfnamefont {D.~A.}\ \bibnamefont {Haber}},\ and\ \bibinfo {author} {\bibfnamefont {S.}~\bibnamefont {Maheswaran}},\ }\bibfield  {title} {\bibinfo {title} {Circulating tumor cell clusters are oligoclonal precursors of breast cancer metastasis},\ }\href {https://doi.org/10.1016/j.cell.2014.07.013} {\bibfield  {journal} {\bibinfo  {journal} {Cell}\ }\textbf {\bibinfo {volume} {158}},\ \bibinfo {pages} {1110} (\bibinfo {year} {2014})}\BibitemShut {NoStop}%
\bibitem [{\citenamefont {Friedl}\ \emph {et~al.}(1995)\citenamefont {Friedl}, \citenamefont {Noble}, \citenamefont {Walton}, \citenamefont {Laird}, \citenamefont {Chauvin}, \citenamefont {Tabah}, \citenamefont {Black},\ and\ \citenamefont {Zänker}}]{Friedl_MigrationOfCoordinatedCellClusters}%
  \BibitemOpen
  \bibfield  {author} {\bibinfo {author} {\bibfnamefont {P.}~\bibnamefont {Friedl}}, \bibinfo {author} {\bibfnamefont {P.~B.}\ \bibnamefont {Noble}}, \bibinfo {author} {\bibfnamefont {P.~A.}\ \bibnamefont {Walton}}, \bibinfo {author} {\bibfnamefont {D.~W.}\ \bibnamefont {Laird}}, \bibinfo {author} {\bibfnamefont {P.~J.}\ \bibnamefont {Chauvin}}, \bibinfo {author} {\bibfnamefont {R.~J.}\ \bibnamefont {Tabah}}, \bibinfo {author} {\bibfnamefont {M.}~\bibnamefont {Black}},\ and\ \bibinfo {author} {\bibfnamefont {K.~S.}\ \bibnamefont {Zänker}},\ }\bibfield  {title} {\bibinfo {title} {{Migration of Coordinated Cell Clusters in Mesenchymal and Epithelial Cancer Explants in Vitro1}},\ }\href@noop {} {\bibfield  {journal} {\bibinfo  {journal} {Cancer Research}\ }\textbf {\bibinfo {volume} {55}},\ \bibinfo {pages} {4557} (\bibinfo {year} {1995})},\ \Eprint {https://arxiv.org/abs/https://aacrjournals.org/cancerres/article-pdf/55/20/4557/2458356/cr0550204557.pdf}
  {https://aacrjournals.org/cancerres/article-pdf/55/20/4557/2458356/cr0550204557.pdf} \BibitemShut {NoStop}%
\bibitem [{\citenamefont {Cheung}\ and\ \citenamefont {Ewald}(2016)}]{Cheung_perspectiveonclustermetastasis}%
  \BibitemOpen
  \bibfield  {author} {\bibinfo {author} {\bibfnamefont {K.~J.}\ \bibnamefont {Cheung}}\ and\ \bibinfo {author} {\bibfnamefont {A.~J.}\ \bibnamefont {Ewald}},\ }\bibfield  {title} {\bibinfo {title} {A collective route to metastasis: Seeding by tumor cell clusters},\ }\href {https://doi.org/10.1126/science.aaf6546} {\bibfield  {journal} {\bibinfo  {journal} {Science}\ }\textbf {\bibinfo {volume} {352}},\ \bibinfo {pages} {167} (\bibinfo {year} {2016})}\BibitemShut {NoStop}%
\bibitem [{\citenamefont {Pi{\~{n}}eiro}\ \emph {et~al.}(2020)\citenamefont {Pi{\~{n}}eiro}, \citenamefont {Mart{\'i}nez-Pena},\ and\ \citenamefont {L{\'o}pez-L{\'o}pez}}]{Pinearo_relevanteofCTCclustersinbreastcancer}%
  \BibitemOpen
  \bibfield  {author} {\bibinfo {author} {\bibfnamefont {R.}~\bibnamefont {Pi{\~{n}}eiro}}, \bibinfo {author} {\bibfnamefont {I.}~\bibnamefont {Mart{\'i}nez-Pena}},\ and\ \bibinfo {author} {\bibfnamefont {R.}~\bibnamefont {L{\'o}pez-L{\'o}pez}},\ }\bibinfo {title} {Relevance of ctc clusters in breast cancer metastasis},\ in\ \href {https://doi.org/10.1007/978-3-030-35805-1_7} {\emph {\bibinfo {booktitle} {Circulating Tumor Cells in Breast Cancer Metastatic Disease}}},\ \bibinfo {editor} {edited by\ \bibinfo {editor} {\bibfnamefont {R.}~\bibnamefont {Pi{\~{n}}eiro}}}\ (\bibinfo  {publisher} {Springer International Publishing},\ \bibinfo {address} {Cham},\ \bibinfo {year} {2020})\ Chap.~\bibinfo {chapter} {7}, pp.\ \bibinfo {pages} {93--115}\BibitemShut {NoStop}%
\bibitem [{\citenamefont {Martínez-Pena}\ \emph {et~al.}(2021)\citenamefont {Martínez-Pena}, \citenamefont {Hurtado}, \citenamefont {Carmona-Ule}, \citenamefont {Abuín}, \citenamefont {Dávila-Ibáñez}, \citenamefont {Sánchez}, \citenamefont {Abal}, \citenamefont {Chaachou}, \citenamefont {Hernández-Losa}, \citenamefont {Cajal}, \citenamefont {López-López},\ and\ \citenamefont {Piñeiro}}]{Matrinex_CTCs_in_zebrafish}%
  \BibitemOpen
  \bibfield  {author} {\bibinfo {author} {\bibfnamefont {I.}~\bibnamefont {Martínez-Pena}}, \bibinfo {author} {\bibfnamefont {P.}~\bibnamefont {Hurtado}}, \bibinfo {author} {\bibfnamefont {N.}~\bibnamefont {Carmona-Ule}}, \bibinfo {author} {\bibfnamefont {C.}~\bibnamefont {Abuín}}, \bibinfo {author} {\bibfnamefont {A.~B.}\ \bibnamefont {Dávila-Ibáñez}}, \bibinfo {author} {\bibfnamefont {L.}~\bibnamefont {Sánchez}}, \bibinfo {author} {\bibfnamefont {M.}~\bibnamefont {Abal}}, \bibinfo {author} {\bibfnamefont {A.}~\bibnamefont {Chaachou}}, \bibinfo {author} {\bibfnamefont {J.}~\bibnamefont {Hernández-Losa}}, \bibinfo {author} {\bibfnamefont {S.~R.~Y.}\ \bibnamefont {Cajal}}, \bibinfo {author} {\bibfnamefont {R.}~\bibnamefont {López-López}},\ and\ \bibinfo {author} {\bibfnamefont {R.}~\bibnamefont {Piñeiro}},\ }\bibfield  {title} {\bibinfo {title} {Dissecting breast cancer circulating tumor cells competence via modelling metastasis in zebrafish},\ }\bibfield  {journal} {\bibinfo  {journal}
  {International Journal of Molecular Sciences}\ }\textbf {\bibinfo {volume} {22}},\ \href {https://doi.org/10.3390/ijms22179279} {10.3390/ijms22179279} (\bibinfo {year} {2021})\BibitemShut {NoStop}%
\bibitem [{\citenamefont {Bocci}\ \emph {et~al.}(2019)\citenamefont {Bocci}, \citenamefont {Kumar~Jolly},\ and\ \citenamefont {Onuchic}}]{Bocci_BiophysicalModel_uncovers_CSD_acrossCancer}%
  \BibitemOpen
  \bibfield  {author} {\bibinfo {author} {\bibfnamefont {F.}~\bibnamefont {Bocci}}, \bibinfo {author} {\bibfnamefont {M.}~\bibnamefont {Kumar~Jolly}},\ and\ \bibinfo {author} {\bibfnamefont {J.~N.}\ \bibnamefont {Onuchic}},\ }\bibfield  {title} {\bibinfo {title} {{A Biophysical Model Uncovers the Size Distribution of Migrating Cell Clusters across Cancer Types}},\ }\href {https://doi.org/10.1158/0008-5472.CAN-19-1726} {\bibfield  {journal} {\bibinfo  {journal} {Cancer Research}\ }\textbf {\bibinfo {volume} {79}},\ \bibinfo {pages} {5527} (\bibinfo {year} {2019})}\BibitemShut {NoStop}%
\bibitem [{\citenamefont {Cheung}\ \emph {et~al.}(2016)\citenamefont {Cheung}, \citenamefont {Padmanaban}, \citenamefont {Silvestri}, \citenamefont {Schipper}, \citenamefont {Cohen}, \citenamefont {Fairchild}, \citenamefont {Gorin}, \citenamefont {Verdone}, \citenamefont {Pienta}, \citenamefont {Bader},\ and\ \citenamefont {Ewald}}]{Cheung_polyclonalbreastcancer_from_collectivedissemination}%
  \BibitemOpen
  \bibfield  {author} {\bibinfo {author} {\bibfnamefont {K.~J.}\ \bibnamefont {Cheung}}, \bibinfo {author} {\bibfnamefont {V.}~\bibnamefont {Padmanaban}}, \bibinfo {author} {\bibfnamefont {V.}~\bibnamefont {Silvestri}}, \bibinfo {author} {\bibfnamefont {K.}~\bibnamefont {Schipper}}, \bibinfo {author} {\bibfnamefont {J.~D.}\ \bibnamefont {Cohen}}, \bibinfo {author} {\bibfnamefont {A.~N.}\ \bibnamefont {Fairchild}}, \bibinfo {author} {\bibfnamefont {M.~A.}\ \bibnamefont {Gorin}}, \bibinfo {author} {\bibfnamefont {J.~E.}\ \bibnamefont {Verdone}}, \bibinfo {author} {\bibfnamefont {K.~J.}\ \bibnamefont {Pienta}}, \bibinfo {author} {\bibfnamefont {J.~S.}\ \bibnamefont {Bader}},\ and\ \bibinfo {author} {\bibfnamefont {A.~J.}\ \bibnamefont {Ewald}},\ }\bibfield  {title} {\bibinfo {title} {Polyclonal breast cancer metastases arise from collective dissemination of keratin 14-expressing tumor cell clusters},\ }\href {https://doi.org/10.1073/pnas.1508541113} {\bibfield  {journal} {\bibinfo  {journal} {Proceedings of the
  National Academy of Sciences}\ }\textbf {\bibinfo {volume} {113}},\ \bibinfo {pages} {E854} (\bibinfo {year} {2016})}\BibitemShut {NoStop}%
\bibitem [{\citenamefont {Kalluri}\ and\ \citenamefont {Weinberg}(2009)}]{Kalluri_basicsofEMT}%
  \BibitemOpen
  \bibfield  {author} {\bibinfo {author} {\bibfnamefont {R.}~\bibnamefont {Kalluri}}\ and\ \bibinfo {author} {\bibfnamefont {R.}~\bibnamefont {Weinberg}},\ }\bibfield  {title} {\bibinfo {title} {The basics of epithelial-mesenchymal transition},\ }\href {https://doi.org/10.1172/JCI39104} {\bibfield  {journal} {\bibinfo  {journal} {The Journal of clinical investigation}\ }\textbf {\bibinfo {volume} {119}},\ \bibinfo {pages} {1420} (\bibinfo {year} {2009})}\BibitemShut {NoStop}%
\bibitem [{\citenamefont {Nieto}\ \emph {et~al.}(2016)\citenamefont {Nieto}, \citenamefont {Huang}, \citenamefont {Jackson},\ and\ \citenamefont {Thiery}}]{Nieto_EMT2016}%
  \BibitemOpen
  \bibfield  {author} {\bibinfo {author} {\bibfnamefont {M.~A.}\ \bibnamefont {Nieto}}, \bibinfo {author} {\bibfnamefont {R.~Y.-J.}\ \bibnamefont {Huang}}, \bibinfo {author} {\bibfnamefont {R.~A.}\ \bibnamefont {Jackson}},\ and\ \bibinfo {author} {\bibfnamefont {J.~P.}\ \bibnamefont {Thiery}},\ }\bibfield  {title} {\bibinfo {title} {Emt: 2016},\ }\href {https://doi.org/10.1016/j.cell.2016.06.028} {\bibfield  {journal} {\bibinfo  {journal} {Cell}\ }\textbf {\bibinfo {volume} {166}},\ \bibinfo {pages} {21} (\bibinfo {year} {2016})}\BibitemShut {NoStop}%
\bibitem [{\citenamefont {Bonnomet}\ \emph {et~al.}(2011)\citenamefont {Bonnomet}, \citenamefont {Syne}, \citenamefont {Feyereisen}, \citenamefont {Thompson}, \citenamefont {Noël}, \citenamefont {Foidart}, \citenamefont {Birembaut}, \citenamefont {Polette},\ and\ \citenamefont {Gilles}}]{Bonnomet_invivo_EMT_in_CTCsbreastcancer}%
  \BibitemOpen
  \bibfield  {author} {\bibinfo {author} {\bibfnamefont {A.}~\bibnamefont {Bonnomet}}, \bibinfo {author} {\bibfnamefont {L.}~\bibnamefont {Syne}}, \bibinfo {author} {\bibfnamefont {E.}~\bibnamefont {Feyereisen}}, \bibinfo {author} {\bibfnamefont {E.}~\bibnamefont {Thompson}}, \bibinfo {author} {\bibfnamefont {A.}~\bibnamefont {Noël}}, \bibinfo {author} {\bibfnamefont {J.-M.}\ \bibnamefont {Foidart}}, \bibinfo {author} {\bibfnamefont {P.}~\bibnamefont {Birembaut}}, \bibinfo {author} {\bibfnamefont {M.}~\bibnamefont {Polette}},\ and\ \bibinfo {author} {\bibfnamefont {C.}~\bibnamefont {Gilles}},\ }\bibfield  {title} {\bibinfo {title} {A dynamic in vivo model of epithelial-to-mesenchymal transitions in circulating tumor cells and metastases of breast cancer},\ }\href {https://doi.org/10.1038/onc.2011.540} {\bibfield  {journal} {\bibinfo  {journal} {Oncogene}\ }\textbf {\bibinfo {volume} {31}},\ \bibinfo {pages} {3741} (\bibinfo {year} {2011})}\BibitemShut {NoStop}%
\bibitem [{\citenamefont {Carey}\ \emph {et~al.}(2013)\citenamefont {Carey}, \citenamefont {Starchenko}, \citenamefont {McGregor},\ and\ \citenamefont {Reinhart-King}}]{Carey_heterogeneous_tumor_exp}%
  \BibitemOpen
  \bibfield  {author} {\bibinfo {author} {\bibfnamefont {S.~P.}\ \bibnamefont {Carey}}, \bibinfo {author} {\bibfnamefont {A.}~\bibnamefont {Starchenko}}, \bibinfo {author} {\bibfnamefont {A.~L.}\ \bibnamefont {McGregor}},\ and\ \bibinfo {author} {\bibfnamefont {C.~A.}\ \bibnamefont {Reinhart-King}},\ }\bibfield  {title} {\bibinfo {title} {Leading malignant cells initiate collective epithelial cell invasion in a three-dimensional heterotypic tumor spheroid model},\ }\href {https://doi.org/10.1007/s10585-013-9565-x} {\bibfield  {journal} {\bibinfo  {journal} {Clinical \& Experimental Metastasis}\ }\textbf {\bibinfo {volume} {30}},\ \bibinfo {pages} {615} (\bibinfo {year} {2013})}\BibitemShut {NoStop}%
\bibitem [{\citenamefont {Jolly}\ \emph {et~al.}(2015)\citenamefont {Jolly}, \citenamefont {Boareto}, \citenamefont {Huang}, \citenamefont {Jia}, \citenamefont {Lu}, \citenamefont {Ben-Jacob}, \citenamefont {Onuchic},\ and\ \citenamefont {Levine}}]{Jolly_ImplicationsOfHybridEMT}%
  \BibitemOpen
  \bibfield  {author} {\bibinfo {author} {\bibfnamefont {M.~K.}\ \bibnamefont {Jolly}}, \bibinfo {author} {\bibfnamefont {M.}~\bibnamefont {Boareto}}, \bibinfo {author} {\bibfnamefont {B.}~\bibnamefont {Huang}}, \bibinfo {author} {\bibfnamefont {D.}~\bibnamefont {Jia}}, \bibinfo {author} {\bibfnamefont {M.}~\bibnamefont {Lu}}, \bibinfo {author} {\bibfnamefont {E.}~\bibnamefont {Ben-Jacob}}, \bibinfo {author} {\bibfnamefont {J.~N.}\ \bibnamefont {Onuchic}},\ and\ \bibinfo {author} {\bibfnamefont {H.}~\bibnamefont {Levine}},\ }\bibfield  {title} {\bibinfo {title} {Implications of the hybrid epithelial/mesenchymal phenotype in metastasis},\ }\href {https://doi.org/10.3389/fonc.2015.00155} {\bibfield  {journal} {\bibinfo  {journal} {Frontiers in Oncology}\ }\textbf {\bibinfo {volume} {5}},\ \bibinfo {pages} {155} (\bibinfo {year} {2015})},\ \bibinfo {note} {eCollection 2015}\BibitemShut {NoStop}%
\bibitem [{\citenamefont {Camley}\ and\ \citenamefont {Rappel}(2017)}]{Camley_PhysicalModels_CollectiveCellMotility}%
  \BibitemOpen
  \bibfield  {author} {\bibinfo {author} {\bibfnamefont {B.~A.}\ \bibnamefont {Camley}}\ and\ \bibinfo {author} {\bibfnamefont {W.-J.}\ \bibnamefont {Rappel}},\ }\bibfield  {title} {\bibinfo {title} {Physical models of collective cell motility: from cell to tissue},\ }\href {https://doi.org/10.1088/1361-6463/aa56fe} {\bibfield  {journal} {\bibinfo  {journal} {Journal of Physics D: Applied Physics}\ }\textbf {\bibinfo {volume} {50}},\ \bibinfo {pages} {113002} (\bibinfo {year} {2017})}\BibitemShut {NoStop}%
\bibitem [{\citenamefont {Alert}\ and\ \citenamefont {Trepat}(2020)}]{Alert_PhysicalModels_of_CollectiveMigration}%
  \BibitemOpen
  \bibfield  {author} {\bibinfo {author} {\bibfnamefont {R.}~\bibnamefont {Alert}}\ and\ \bibinfo {author} {\bibfnamefont {X.}~\bibnamefont {Trepat}},\ }\bibfield  {title} {\bibinfo {title} {Physical models of collective cell migration},\ }\href {https://doi.org/10.1146/annurev-conmatphys-031218-013516} {\bibfield  {journal} {\bibinfo  {journal} {Annual Review of Condensed Matter Physics}\ }\textbf {\bibinfo {volume} {11}},\ \bibinfo {pages} {77} (\bibinfo {year} {2020})}\BibitemShut {NoStop}%
\bibitem [{\citenamefont {Lv}\ \emph {et~al.}(2021)\citenamefont {Lv}, \citenamefont {Chen}, \citenamefont {Guan}, \citenamefont {Góźdź}, \citenamefont {Feng},\ and\ \citenamefont {Li}}]{Lv_collectivemigration_in_epithelial_cancerousmonolayer}%
  \BibitemOpen
  \bibfield  {author} {\bibinfo {author} {\bibfnamefont {J.-Q.}\ \bibnamefont {Lv}}, \bibinfo {author} {\bibfnamefont {P.-C.}\ \bibnamefont {Chen}}, \bibinfo {author} {\bibfnamefont {L.-Y.}\ \bibnamefont {Guan}}, \bibinfo {author} {\bibfnamefont {W.~T.}\ \bibnamefont {Góźdź}}, \bibinfo {author} {\bibfnamefont {X.-Q.}\ \bibnamefont {Feng}},\ and\ \bibinfo {author} {\bibfnamefont {B.}~\bibnamefont {Li}},\ }\bibfield  {title} {\bibinfo {title} {Collective migrations in an epithelial–cancerous cell monolayer},\ }\href {https://doi.org/10.1007/s10409-021-01083-1} {\bibfield  {journal} {\bibinfo  {journal} {Acta Mechanica Sinica}\ }\textbf {\bibinfo {volume} {37}},\ \bibinfo {pages} {773–784} (\bibinfo {year} {2021})}\BibitemShut {NoStop}%
\bibitem [{\citenamefont {Nakajima}\ and\ \citenamefont {Ishihara}(2011)}]{Nakajima_kinetics_CPM_revised}%
  \BibitemOpen
  \bibfield  {author} {\bibinfo {author} {\bibfnamefont {A.}~\bibnamefont {Nakajima}}\ and\ \bibinfo {author} {\bibfnamefont {S.}~\bibnamefont {Ishihara}},\ }\bibfield  {title} {\bibinfo {title} {Kinetics of the cellular potts model revisited},\ }\href {https://doi.org/10.1088/1367-2630/13/3/033035} {\bibfield  {journal} {\bibinfo  {journal} {New Journal of Physics}\ }\textbf {\bibinfo {volume} {13}},\ \bibinfo {pages} {033035} (\bibinfo {year} {2011})}\BibitemShut {NoStop}%
\bibitem [{\citenamefont {Beatrici}\ \emph {et~al.}(2017)\citenamefont {Beatrici}, \citenamefont {de~Almeida},\ and\ \citenamefont {Brunnet}}]{Beatrici_meanclusterapproach_cellsorting}%
  \BibitemOpen
  \bibfield  {author} {\bibinfo {author} {\bibfnamefont {C.~P.}\ \bibnamefont {Beatrici}}, \bibinfo {author} {\bibfnamefont {R.~M.~C.}\ \bibnamefont {de~Almeida}},\ and\ \bibinfo {author} {\bibfnamefont {L.~G.}\ \bibnamefont {Brunnet}},\ }\bibfield  {title} {\bibinfo {title} {Mean-cluster approach indicates cell sorting time scales are determined by collective dynamics},\ }\href {https://doi.org/10.1103/PhysRevE.95.032402} {\bibfield  {journal} {\bibinfo  {journal} {Physical Review E}\ }\textbf {\bibinfo {volume} {95}},\ \bibinfo {pages} {032402} (\bibinfo {year} {2017})}\BibitemShut {NoStop}%
\bibitem [{\citenamefont {Hallou}\ \emph {et~al.}(2017)\citenamefont {Hallou}, \citenamefont {Jennings},\ and\ \citenamefont {Kabla}}]{Hallou_tumorheterogeneitypromotescollectiveinvasion}%
  \BibitemOpen
  \bibfield  {author} {\bibinfo {author} {\bibfnamefont {A.}~\bibnamefont {Hallou}}, \bibinfo {author} {\bibfnamefont {J.}~\bibnamefont {Jennings}},\ and\ \bibinfo {author} {\bibfnamefont {A.~J.}\ \bibnamefont {Kabla}},\ }\bibfield  {title} {\bibinfo {title} {Tumour heterogeneity promotes collective invasion and cancer metastatic dissemination},\ }\href {https://doi.org/10.1098/rsos.161007} {\bibfield  {journal} {\bibinfo  {journal} {Royal Society Open Science}\ }\textbf {\bibinfo {volume} {4}},\ \bibinfo {pages} {161007} (\bibinfo {year} {2017})}\BibitemShut {NoStop}%
\bibitem [{\citenamefont {Kabla}(2012)}]{Kabla_collectivecellmigration}%
  \BibitemOpen
  \bibfield  {author} {\bibinfo {author} {\bibfnamefont {A.~J.}\ \bibnamefont {Kabla}},\ }\bibfield  {title} {\bibinfo {title} {Collective cell migration: Leadership, invasion and segregation},\ }\href {https://doi.org/10.1098/rsif.2012.0448} {\bibfield  {journal} {\bibinfo  {journal} {Journal of the Royal Society Interface}\ }\textbf {\bibinfo {volume} {9}},\ \bibinfo {pages} {3268} (\bibinfo {year} {2012})}\BibitemShut {NoStop}%
\bibitem [{\citenamefont {Belmonte}\ \emph {et~al.}(2008)\citenamefont {Belmonte}, \citenamefont {Thomas}, \citenamefont {Brunnet}, \citenamefont {de~Almeida},\ and\ \citenamefont {Chat\'e}}]{Belmonte_SPPmodelforcellsorting}%
  \BibitemOpen
  \bibfield  {author} {\bibinfo {author} {\bibfnamefont {J.~M.}\ \bibnamefont {Belmonte}}, \bibinfo {author} {\bibfnamefont {G.~L.}\ \bibnamefont {Thomas}}, \bibinfo {author} {\bibfnamefont {L.~G.}\ \bibnamefont {Brunnet}}, \bibinfo {author} {\bibfnamefont {R.~M.~C.}\ \bibnamefont {de~Almeida}},\ and\ \bibinfo {author} {\bibfnamefont {H.}~\bibnamefont {Chat\'e}},\ }\bibfield  {title} {\bibinfo {title} {Self-propelled particle model for cell-sorting phenomena},\ }\href {https://doi.org/10.1103/PhysRevLett.100.248702} {\bibfield  {journal} {\bibinfo  {journal} {Phys. Rev. Lett.}\ }\textbf {\bibinfo {volume} {100}},\ \bibinfo {pages} {248702} (\bibinfo {year} {2008})}\BibitemShut {NoStop}%
\bibitem [{\citenamefont {Martín-Gómez}\ \emph {et~al.}(2018)\citenamefont {Martín-Gómez}, \citenamefont {Levis}, \citenamefont {Díaz-Guilera},\ and\ \citenamefont {Pagonabarraga}}]{MartinGomez_CollectiveMotion_PolarAlignment}%
  \BibitemOpen
  \bibfield  {author} {\bibinfo {author} {\bibfnamefont {A.}~\bibnamefont {Martín-Gómez}}, \bibinfo {author} {\bibfnamefont {D.}~\bibnamefont {Levis}}, \bibinfo {author} {\bibfnamefont {A.}~\bibnamefont {Díaz-Guilera}},\ and\ \bibinfo {author} {\bibfnamefont {I.}~\bibnamefont {Pagonabarraga}},\ }\bibfield  {title} {\bibinfo {title} {Collective motion of active brownian particles with polar alignment},\ }\href {https://doi.org/10.1039/C8SM00020D} {\bibfield  {journal} {\bibinfo  {journal} {Soft Matter}\ }\textbf {\bibinfo {volume} {14}},\ \bibinfo {pages} {2610} (\bibinfo {year} {2018})}\BibitemShut {NoStop}%
\bibitem [{\citenamefont {L{\aa}ng}\ \emph {et~al.}(2018)\citenamefont {L{\aa}ng}, \citenamefont {Po{\l}e{\'c}}, \citenamefont {L{\aa}ng}, \citenamefont {Valk}, \citenamefont {Blicher}, \citenamefont {Rowe}, \citenamefont {T{\o}nseth}, \citenamefont {Jackson}, \citenamefont {Utheim}, \citenamefont {Janssen} \emph {et~al.}}]{laang2018coordinated}%
  \BibitemOpen
  \bibfield  {author} {\bibinfo {author} {\bibfnamefont {E.}~\bibnamefont {L{\aa}ng}}, \bibinfo {author} {\bibfnamefont {A.}~\bibnamefont {Po{\l}e{\'c}}}, \bibinfo {author} {\bibfnamefont {A.}~\bibnamefont {L{\aa}ng}}, \bibinfo {author} {\bibfnamefont {M.}~\bibnamefont {Valk}}, \bibinfo {author} {\bibfnamefont {P.}~\bibnamefont {Blicher}}, \bibinfo {author} {\bibfnamefont {A.~D.}\ \bibnamefont {Rowe}}, \bibinfo {author} {\bibfnamefont {K.~A.}\ \bibnamefont {T{\o}nseth}}, \bibinfo {author} {\bibfnamefont {C.~J.}\ \bibnamefont {Jackson}}, \bibinfo {author} {\bibfnamefont {T.~P.}\ \bibnamefont {Utheim}}, \bibinfo {author} {\bibfnamefont {L.~M.}\ \bibnamefont {Janssen}}, \emph {et~al.},\ }\bibfield  {title} {\bibinfo {title} {Coordinated collective migration and asymmetric cell division in confluent human keratinocytes without wounding},\ }\href@noop {} {\bibfield  {journal} {\bibinfo  {journal} {Nature communications}\ }\textbf {\bibinfo {volume} {9}},\ \bibinfo {pages} {3665} (\bibinfo {year}
  {2018})}\BibitemShut {NoStop}%
\bibitem [{\citenamefont {Sepúlveda}\ \emph {et~al.}(2013)\citenamefont {Sepúlveda}, \citenamefont {Petitjean}, \citenamefont {Cochet}, \citenamefont {Grasland-Mongrain}, \citenamefont {Silberzan},\ and\ \citenamefont {Hakim}}]{Sepulveda_collectivecellmotion}%
  \BibitemOpen
  \bibfield  {author} {\bibinfo {author} {\bibfnamefont {N.}~\bibnamefont {Sepúlveda}}, \bibinfo {author} {\bibfnamefont {L.}~\bibnamefont {Petitjean}}, \bibinfo {author} {\bibfnamefont {O.}~\bibnamefont {Cochet}}, \bibinfo {author} {\bibfnamefont {E.}~\bibnamefont {Grasland-Mongrain}}, \bibinfo {author} {\bibfnamefont {P.}~\bibnamefont {Silberzan}},\ and\ \bibinfo {author} {\bibfnamefont {V.}~\bibnamefont {Hakim}},\ }\bibfield  {title} {\bibinfo {title} {Collective cell motion in an epithelial sheet can be quantitatively described by a stochastic interacting particle model},\ }\href {https://doi.org/10.1371/journal.pcbi.1002944} {\bibfield  {journal} {\bibinfo  {journal} {PLoS Computational Biology}\ }\textbf {\bibinfo {volume} {9}},\ \bibinfo {pages} {e1002944} (\bibinfo {year} {2013})}\BibitemShut {NoStop}%
\bibitem [{\citenamefont {Rappel}\ \emph {et~al.}(1999)\citenamefont {Rappel}, \citenamefont {Nicol}, \citenamefont {Sarkissian}, \citenamefont {Levine},\ and\ \citenamefont {Loomis}}]{Rappel_selforganizedvortexstate}%
  \BibitemOpen
  \bibfield  {author} {\bibinfo {author} {\bibfnamefont {W.-J.}\ \bibnamefont {Rappel}}, \bibinfo {author} {\bibfnamefont {A.}~\bibnamefont {Nicol}}, \bibinfo {author} {\bibfnamefont {A.}~\bibnamefont {Sarkissian}}, \bibinfo {author} {\bibfnamefont {H.}~\bibnamefont {Levine}},\ and\ \bibinfo {author} {\bibfnamefont {W.~F.}\ \bibnamefont {Loomis}},\ }\bibfield  {title} {\bibinfo {title} {Self-organized vortex state in two-dimensional dictyostelium dynamics},\ }\href {https://doi.org/10.1103/PhysRevLett.83.1247} {\bibfield  {journal} {\bibinfo  {journal} {Physical Review Letters}\ }\textbf {\bibinfo {volume} {83}},\ \bibinfo {pages} {1247} (\bibinfo {year} {1999})}\BibitemShut {NoStop}%
\bibitem [{\citenamefont {Rubenstein}\ and\ \citenamefont {Kaufman}(2008)}]{rubenstein2008role}%
  \BibitemOpen
  \bibfield  {author} {\bibinfo {author} {\bibfnamefont {B.~M.}\ \bibnamefont {Rubenstein}}\ and\ \bibinfo {author} {\bibfnamefont {L.~J.}\ \bibnamefont {Kaufman}},\ }\bibfield  {title} {\bibinfo {title} {The role of extracellular matrix in glioma invasion: a cellular potts model approach},\ }\href@noop {} {\bibfield  {journal} {\bibinfo  {journal} {Biophysical journal}\ }\textbf {\bibinfo {volume} {95}},\ \bibinfo {pages} {5661} (\bibinfo {year} {2008})}\BibitemShut {NoStop}%
\bibitem [{\citenamefont {Mar{\'e}e}\ \emph {et~al.}(2007)\citenamefont {Mar{\'e}e}, \citenamefont {Grieneisen},\ and\ \citenamefont {Hogeweg}}]{Maree_CPM_book_Hogeweg}%
  \BibitemOpen
  \bibfield  {author} {\bibinfo {author} {\bibfnamefont {A.~F.~M.}\ \bibnamefont {Mar{\'e}e}}, \bibinfo {author} {\bibfnamefont {V.~A.}\ \bibnamefont {Grieneisen}},\ and\ \bibinfo {author} {\bibfnamefont {P.}~\bibnamefont {Hogeweg}},\ }\bibinfo {title} {The cellular potts model and biophysical properties of cells, tissues and morphogenesis},\ in\ \href {https://doi.org/10.1007/978-3-7643-8123-3_5} {\emph {\bibinfo {booktitle} {Single-Cell-Based Models in Biology and Medicine}}}\ (\bibinfo  {publisher} {Birkh{\"a}user Basel},\ \bibinfo {address} {Basel},\ \bibinfo {year} {2007})\ Chap.~\bibinfo {chapter} {2}, pp.\ \bibinfo {pages} {107--136}\BibitemShut {NoStop}%
\bibitem [{\citenamefont {Guisoni}\ \emph {et~al.}(2018)\citenamefont {Guisoni}, \citenamefont {Mazzitello},\ and\ \citenamefont {Diambra}}]{Guisoni_ModellingActiveCellMovement}%
  \BibitemOpen
  \bibfield  {author} {\bibinfo {author} {\bibfnamefont {N.}~\bibnamefont {Guisoni}}, \bibinfo {author} {\bibfnamefont {K.~I.}\ \bibnamefont {Mazzitello}},\ and\ \bibinfo {author} {\bibfnamefont {L.}~\bibnamefont {Diambra}},\ }\bibfield  {title} {\bibinfo {title} {Modeling active cell movement with the potts model},\ }\bibfield  {journal} {\bibinfo  {journal} {Frontiers in Physics}\ }\textbf {\bibinfo {volume} {6}},\ \href {https://doi.org/10.3389/fphy.2018.00061} {10.3389/fphy.2018.00061} (\bibinfo {year} {2018})\BibitemShut {NoStop}%
\bibitem [{\citenamefont {Scianna}\ \emph {et~al.}(2013)\citenamefont {Scianna}, \citenamefont {Preziosi},\ and\ \citenamefont {Wolf}}]{Scianna_ECM}%
  \BibitemOpen
  \bibfield  {author} {\bibinfo {author} {\bibfnamefont {M.}~\bibnamefont {Scianna}}, \bibinfo {author} {\bibfnamefont {L.}~\bibnamefont {Preziosi}},\ and\ \bibinfo {author} {\bibfnamefont {K.}~\bibnamefont {Wolf}},\ }\bibfield  {title} {\bibinfo {title} {A cellular potts model simulating cell migration on and in matrix environments},\ }\href {https://doi.org/10.3934/mbe.2013.10.235} {\bibfield  {journal} {\bibinfo  {journal} {Mathematical Biosciences and Engineering}\ }\textbf {\bibinfo {volume} {10}},\ \bibinfo {pages} {235} (\bibinfo {year} {2013})}\BibitemShut {NoStop}%
\bibitem [{\citenamefont {Matsushita}(2017)}]{Katsuyoshi_CellAlignment_PolarizedAdhesion}%
  \BibitemOpen
  \bibfield  {author} {\bibinfo {author} {\bibfnamefont {K.}~\bibnamefont {Matsushita}},\ }\bibfield  {title} {\bibinfo {title} {Cell-alignment patterns in the collective migration of cells with polarized adhesion},\ }\href {https://doi.org/10.1103/PhysRevE.95.032415} {\bibfield  {journal} {\bibinfo  {journal} {Phys. Rev. E}\ }\textbf {\bibinfo {volume} {95}},\ \bibinfo {pages} {032415} (\bibinfo {year} {2017})}\BibitemShut {NoStop}%
\bibitem [{\citenamefont {Devanny}\ \emph {et~al.}(2023)\citenamefont {Devanny}, \citenamefont {Lee}, \citenamefont {Kampman},\ and\ \citenamefont {Kaufman}}]{devanny2023signatures}%
  \BibitemOpen
  \bibfield  {author} {\bibinfo {author} {\bibfnamefont {A.~J.}\ \bibnamefont {Devanny}}, \bibinfo {author} {\bibfnamefont {D.~J.}\ \bibnamefont {Lee}}, \bibinfo {author} {\bibfnamefont {L.}~\bibnamefont {Kampman}},\ and\ \bibinfo {author} {\bibfnamefont {L.~J.}\ \bibnamefont {Kaufman}},\ }\bibfield  {title} {\bibinfo {title} {Signatures of jamming in the cellular potts model},\ }\href@noop {} {\bibfield  {journal} {\bibinfo  {journal} {bioRxiv}\ ,\ \bibinfo {pages} {2023}} (\bibinfo {year} {2023})}\BibitemShut {NoStop}%
\bibitem [{\citenamefont {Vicsek}\ \emph {et~al.}(1995)\citenamefont {Vicsek}, \citenamefont {Czir\'ok}, \citenamefont {Ben-Jacob}, \citenamefont {Cohen},\ and\ \citenamefont {Shochet}}]{Vicsek_NovelTypePhaseTransition}%
  \BibitemOpen
  \bibfield  {author} {\bibinfo {author} {\bibfnamefont {T.}~\bibnamefont {Vicsek}}, \bibinfo {author} {\bibfnamefont {A.}~\bibnamefont {Czir\'ok}}, \bibinfo {author} {\bibfnamefont {E.}~\bibnamefont {Ben-Jacob}}, \bibinfo {author} {\bibfnamefont {I.}~\bibnamefont {Cohen}},\ and\ \bibinfo {author} {\bibfnamefont {O.}~\bibnamefont {Shochet}},\ }\bibfield  {title} {\bibinfo {title} {Novel type of phase transition in a system of self-driven particles},\ }\href {https://doi.org/10.1103/PhysRevLett.75.1226} {\bibfield  {journal} {\bibinfo  {journal} {Physics Review Letters}\ }\textbf {\bibinfo {volume} {75}},\ \bibinfo {pages} {1226} (\bibinfo {year} {1995})}\BibitemShut {NoStop}%
\bibitem [{\citenamefont {Debets}\ \emph {et~al.}(2021)\citenamefont {Debets}, \citenamefont {Janssen},\ and\ \citenamefont {Storm}}]{Debets_EnhancedPersistence}%
  \BibitemOpen
  \bibfield  {author} {\bibinfo {author} {\bibfnamefont {V.~E.}\ \bibnamefont {Debets}}, \bibinfo {author} {\bibfnamefont {L.~M.~C.}\ \bibnamefont {Janssen}},\ and\ \bibinfo {author} {\bibfnamefont {C.}~\bibnamefont {Storm}},\ }\bibfield  {title} {\bibinfo {title} {Enhanced persistence and collective migration in cooperatively aligning cell clusters},\ }\href {https://doi.org/10.1016/j.bpj.2021.02.014} {\bibfield  {journal} {\bibinfo  {journal} {Biophysical Journal}\ }\textbf {\bibinfo {volume} {120}},\ \bibinfo {pages} {1483} (\bibinfo {year} {2021})}\BibitemShut {NoStop}%
\bibitem [{\citenamefont {Li}\ \emph {et~al.}(2008)\citenamefont {Li}, \citenamefont {Nørrelykke},\ and\ \citenamefont {Cox}}]{Li_persistentrandommotion_absenceofsignals}%
  \BibitemOpen
  \bibfield  {author} {\bibinfo {author} {\bibfnamefont {L.}~\bibnamefont {Li}}, \bibinfo {author} {\bibfnamefont {S.~F.}\ \bibnamefont {Nørrelykke}},\ and\ \bibinfo {author} {\bibfnamefont {E.~C.}\ \bibnamefont {Cox}},\ }\bibfield  {title} {\bibinfo {title} {Persistent cell motion in the absence of external signals: A search strategy for eukaryotic cells},\ }\href {https://doi.org/10.1371/journal.pone.0002093} {\bibfield  {journal} {\bibinfo  {journal} {PLoS One}\ }\textbf {\bibinfo {volume} {3}},\ \bibinfo {pages} {s2093} (\bibinfo {year} {2008})}\BibitemShut {NoStop}%
\bibitem [{\citenamefont {Swat}\ \emph {et~al.}(2012)\citenamefont {Swat}, \citenamefont {Thomas}, \citenamefont {Belmonte}, \citenamefont {Shirinifard}, \citenamefont {Hmeljak},\ and\ \citenamefont {Glazier}}]{Swat_CC3D}%
  \BibitemOpen
  \bibfield  {author} {\bibinfo {author} {\bibfnamefont {M.~H.}\ \bibnamefont {Swat}}, \bibinfo {author} {\bibfnamefont {G.~L.}\ \bibnamefont {Thomas}}, \bibinfo {author} {\bibfnamefont {J.~M.}\ \bibnamefont {Belmonte}}, \bibinfo {author} {\bibfnamefont {A.}~\bibnamefont {Shirinifard}}, \bibinfo {author} {\bibfnamefont {D.}~\bibnamefont {Hmeljak}},\ and\ \bibinfo {author} {\bibfnamefont {J.~A.}\ \bibnamefont {Glazier}},\ }\bibfield  {title} {\bibinfo {title} {Chapter 13 - multi-scale modeling of tissues using compucell3d},\ }in\ \href {https://doi.org/10.1016/B978-0-12-388403-9.00013-8} {\emph {\bibinfo {booktitle} {Computational Methods in Cell Biology}}},\ \bibinfo {series} {Methods in Cell Biology}, Vol.\ \bibinfo {volume} {110},\ \bibinfo {editor} {edited by\ \bibinfo {editor} {\bibfnamefont {A.~R.}\ \bibnamefont {Asthigiri}}\ and\ \bibinfo {editor} {\bibfnamefont {A.~P.}\ \bibnamefont {Arkin}}}\ (\bibinfo  {publisher} {Academic Press},\ \bibinfo {year} {2012})\ pp.\ \bibinfo {pages}
  {325--366}\BibitemShut {NoStop}%
\bibitem [{\citenamefont {Graner}\ and\ \citenamefont {Glazier}(1992)}]{Graner_CellSorting}%
  \BibitemOpen
  \bibfield  {author} {\bibinfo {author} {\bibfnamefont {F.}~\bibnamefont {Graner}}\ and\ \bibinfo {author} {\bibfnamefont {J.~A.}\ \bibnamefont {Glazier}},\ }\bibfield  {title} {\bibinfo {title} {Simulation of biological cell sorting using a two-dimensional extended potts model},\ }\href {https://doi.org/10.1103/PhysRevLett.69.2013} {\bibfield  {journal} {\bibinfo  {journal} {Physics Review Letters}\ }\textbf {\bibinfo {volume} {69}},\ \bibinfo {pages} {2013} (\bibinfo {year} {1992})}\BibitemShut {NoStop}%
\bibitem [{\citenamefont {Glazier}\ and\ \citenamefont {Graner}(1993)}]{Glazier_Simulation_of_DAdrivenrearrangement}%
  \BibitemOpen
  \bibfield  {author} {\bibinfo {author} {\bibfnamefont {J.~A.}\ \bibnamefont {Glazier}}\ and\ \bibinfo {author} {\bibfnamefont {F.}~\bibnamefont {Graner}},\ }\bibfield  {title} {\bibinfo {title} {Simulation of the differential adhesion driven rearrangement of biological cells},\ }\href {https://doi.org/10.1103/PhysRevE.47.2128} {\bibfield  {journal} {\bibinfo  {journal} {Physical Review E}\ }\textbf {\bibinfo {volume} {47}},\ \bibinfo {pages} {2128} (\bibinfo {year} {1993})}\BibitemShut {NoStop}%
\bibitem [{\citenamefont {Hirashima}\ \emph {et~al.}(2017)\citenamefont {Hirashima}, \citenamefont {Rens},\ and\ \citenamefont {Merks}}]{Hirashima_CPMreview_morphogenesis}%
  \BibitemOpen
  \bibfield  {author} {\bibinfo {author} {\bibfnamefont {T.}~\bibnamefont {Hirashima}}, \bibinfo {author} {\bibfnamefont {E.~G.}\ \bibnamefont {Rens}},\ and\ \bibinfo {author} {\bibfnamefont {R.~M.~H.}\ \bibnamefont {Merks}},\ }\bibfield  {title} {\bibinfo {title} {Cellular potts modeling of complex multicellular behaviors in tissue morphogenesis},\ }\href {https://doi.org/10.1111/dgd.12358} {\bibfield  {journal} {\bibinfo  {journal} {Development, Growth \& Differentiation}\ }\textbf {\bibinfo {volume} {59}},\ \bibinfo {pages} {329} (\bibinfo {year} {2017})}\BibitemShut {NoStop}%
\bibitem [{\citenamefont {Scianna}\ and\ \citenamefont {Preziosi}(2013)}]{scianna2013_book}%
  \BibitemOpen
  \bibfield  {author} {\bibinfo {author} {\bibfnamefont {M.}~\bibnamefont {Scianna}}\ and\ \bibinfo {author} {\bibfnamefont {L.}~\bibnamefont {Preziosi}},\ }\href@noop {} {\emph {\bibinfo {title} {Cellular potts models: multiscale extensions and biological applications}}}\ (\bibinfo  {publisher} {Chapman Hall/CRC Press},\ \bibinfo {year} {2013})\BibitemShut {NoStop}%
\bibitem [{\citenamefont {Szabó}\ and\ \citenamefont {Merks}(2013)}]{Szabo_review_CPM_tumor}%
  \BibitemOpen
  \bibfield  {author} {\bibinfo {author} {\bibfnamefont {A.}~\bibnamefont {Szabó}}\ and\ \bibinfo {author} {\bibfnamefont {R.~M.}\ \bibnamefont {Merks}},\ }\bibfield  {title} {\bibinfo {title} {Cellular potts modeling of tumor growth, tumor invasion, and tumor evolution},\ }\bibfield  {journal} {\bibinfo  {journal} {Frontiers in Oncology}\ }\textbf {\bibinfo {volume} {3}},\ \href {https://doi.org/10.3389/fonc.2013.00087} {10.3389/fonc.2013.00087} (\bibinfo {year} {2013})\BibitemShut {NoStop}%
\bibitem [{\citenamefont {Frenkel}\ and\ \citenamefont {Smit}(2002)}]{Frenkel_ChapterMonteCarlo}%
  \BibitemOpen
  \bibfield  {author} {\bibinfo {author} {\bibfnamefont {D.}~\bibnamefont {Frenkel}}\ and\ \bibinfo {author} {\bibfnamefont {B.}~\bibnamefont {Smit}},\ }\bibfield  {title} {\bibinfo {title} {Chapter 3 - monte carlo simulations},\ }in\ \href {https://doi.org/10.1016/B978-012267351-1/50005-5} {\emph {\bibinfo {booktitle} {Understanding Molecular Simulation (Second Edition)}}},\ \bibinfo {editor} {edited by\ \bibinfo {editor} {\bibfnamefont {D.}~\bibnamefont {Frenkel}}\ and\ \bibinfo {editor} {\bibfnamefont {B.}~\bibnamefont {Smit}}}\ (\bibinfo  {publisher} {Academic Press},\ \bibinfo {address} {San Diego},\ \bibinfo {year} {2002})\ \bibinfo {edition} {second edition}\ ed.,\ pp.\ \bibinfo {pages} {23--61}\BibitemShut {NoStop}%
\bibitem [{\citenamefont {Metropolis}\ \emph {et~al.}(1953)\citenamefont {Metropolis}, \citenamefont {Rosenbluth}, \citenamefont {Rosenbluth}, \citenamefont {Teller},\ and\ \citenamefont {Teller}}]{Metropolis_MCMetropolis}%
  \BibitemOpen
  \bibfield  {author} {\bibinfo {author} {\bibfnamefont {N.}~\bibnamefont {Metropolis}}, \bibinfo {author} {\bibfnamefont {A.~W.}\ \bibnamefont {Rosenbluth}}, \bibinfo {author} {\bibfnamefont {M.~N.}\ \bibnamefont {Rosenbluth}}, \bibinfo {author} {\bibfnamefont {A.~H.}\ \bibnamefont {Teller}},\ and\ \bibinfo {author} {\bibfnamefont {E.}~\bibnamefont {Teller}},\ }\bibfield  {title} {\bibinfo {title} {Equation of state calculations by fast computing machines},\ }\href {https://doi.org/10.1063/1.1699114} {\bibfield  {journal} {\bibinfo  {journal} {The Journal of Chemical Physics}\ }\textbf {\bibinfo {volume} {21}},\ \bibinfo {pages} {1087} (\bibinfo {year} {1953})}\BibitemShut {NoStop}%
\bibitem [{\citenamefont {Kim}\ \emph {et~al.}(2020)\citenamefont {Kim}, \citenamefont {Pegoraro}, \citenamefont {Das}, \citenamefont {Koehler}, \citenamefont {Ujwary}, \citenamefont {Lan}, \citenamefont {Mitchel}, \citenamefont {Atia}, \citenamefont {He}, \citenamefont {Wang}, \citenamefont {Bi}, \citenamefont {Zaman}, \citenamefont {Park}, \citenamefont {Butler}, \citenamefont {Lee}, \citenamefont {Starr},\ and\ \citenamefont {Fredberg}}]{Kim_UnjammingMigration_BreastCancerCells}%
  \BibitemOpen
  \bibfield  {author} {\bibinfo {author} {\bibfnamefont {J.~H.}\ \bibnamefont {Kim}}, \bibinfo {author} {\bibfnamefont {A.~F.}\ \bibnamefont {Pegoraro}}, \bibinfo {author} {\bibfnamefont {A.}~\bibnamefont {Das}}, \bibinfo {author} {\bibfnamefont {S.~A.}\ \bibnamefont {Koehler}}, \bibinfo {author} {\bibfnamefont {S.~A.}\ \bibnamefont {Ujwary}}, \bibinfo {author} {\bibfnamefont {B.}~\bibnamefont {Lan}}, \bibinfo {author} {\bibfnamefont {J.~A.}\ \bibnamefont {Mitchel}}, \bibinfo {author} {\bibfnamefont {L.}~\bibnamefont {Atia}}, \bibinfo {author} {\bibfnamefont {S.}~\bibnamefont {He}}, \bibinfo {author} {\bibfnamefont {K.}~\bibnamefont {Wang}}, \bibinfo {author} {\bibfnamefont {D.}~\bibnamefont {Bi}}, \bibinfo {author} {\bibfnamefont {M.~H.}\ \bibnamefont {Zaman}}, \bibinfo {author} {\bibfnamefont {J.-A.}\ \bibnamefont {Park}}, \bibinfo {author} {\bibfnamefont {J.~P.}\ \bibnamefont {Butler}}, \bibinfo {author} {\bibfnamefont {K.~H.}\ \bibnamefont {Lee}}, \bibinfo {author} {\bibfnamefont {J.~R.}\ \bibnamefont
  {Starr}},\ and\ \bibinfo {author} {\bibfnamefont {J.~J.}\ \bibnamefont {Fredberg}},\ }\bibfield  {title} {\bibinfo {title} {Unjamming and collective migration in mcf10a breast cancer cell lines},\ }\href {https://doi.org/10.1016/j.bbrc.2019.10.188} {\bibfield  {journal} {\bibinfo  {journal} {Biochemical and Biophysical Research Communications}\ }\textbf {\bibinfo {volume} {521}},\ \bibinfo {pages} {706} (\bibinfo {year} {2020})}\BibitemShut {NoStop}%
\bibitem [{\citenamefont {West}\ \emph {et~al.}(2017)\citenamefont {West}, \citenamefont {Wullkopf}, \citenamefont {Christensen}, \citenamefont {Leijnse}, \citenamefont {Tarp}, \citenamefont {Mathiesen}, \citenamefont {Erler},\ and\ \citenamefont {Oddershede}}]{West_DynamicsCancerousTissue}%
  \BibitemOpen
  \bibfield  {author} {\bibinfo {author} {\bibfnamefont {A.~K.~V.}\ \bibnamefont {West}}, \bibinfo {author} {\bibfnamefont {L.}~\bibnamefont {Wullkopf}}, \bibinfo {author} {\bibfnamefont {A.}~\bibnamefont {Christensen}}, \bibinfo {author} {\bibfnamefont {N.}~\bibnamefont {Leijnse}}, \bibinfo {author} {\bibfnamefont {J.~M.}\ \bibnamefont {Tarp}}, \bibinfo {author} {\bibfnamefont {J.}~\bibnamefont {Mathiesen}}, \bibinfo {author} {\bibfnamefont {J.~T.}\ \bibnamefont {Erler}},\ and\ \bibinfo {author} {\bibfnamefont {L.~B.}\ \bibnamefont {Oddershede}},\ }\bibfield  {title} {\bibinfo {title} {Dynamics of cancerous tissue correlates with invasiveness},\ }\bibfield  {journal} {\bibinfo  {journal} {Scientific Reports}\ }\textbf {\bibinfo {volume} {7}},\ \href {https://doi.org/10.1038/srep43800} {10.1038/srep43800} (\bibinfo {year} {2017})\BibitemShut {NoStop}%
\bibitem [{\citenamefont {Prasanna}\ \emph {et~al.}(2024)\citenamefont {Prasanna}, \citenamefont {Jolly},\ and\ \citenamefont {Bhat}}]{prasanna2024spatial}%
  \BibitemOpen
  \bibfield  {author} {\bibinfo {author} {\bibfnamefont {C.~V.~S.}\ \bibnamefont {Prasanna}}, \bibinfo {author} {\bibfnamefont {M.~K.}\ \bibnamefont {Jolly}},\ and\ \bibinfo {author} {\bibfnamefont {R.}~\bibnamefont {Bhat}},\ }\bibfield  {title} {\bibinfo {title} {Spatial heterogeneity in tumor adhesion qualifies collective cell invasion},\ }\href@noop {} {\bibfield  {journal} {\bibinfo  {journal} {Biophysical Journal}\ } (\bibinfo {year} {2024})}\BibitemShut {NoStop}%
\bibitem [{\citenamefont {Sandersius}\ \emph {et~al.}(2011)\citenamefont {Sandersius}, \citenamefont {Weijer},\ and\ \citenamefont {Newman}}]{Sandersius_EmergentCellAndTissueDynamics}%
  \BibitemOpen
  \bibfield  {author} {\bibinfo {author} {\bibfnamefont {S.~A.}\ \bibnamefont {Sandersius}}, \bibinfo {author} {\bibfnamefont {C.~J.}\ \bibnamefont {Weijer}},\ and\ \bibinfo {author} {\bibfnamefont {T.~J.}\ \bibnamefont {Newman}},\ }\bibfield  {title} {\bibinfo {title} {Emergent cell and tissue dynamics from subcellular modeling of active biomechanical processes},\ }\href {https://doi.org/10.1088/1478-3975/8/4/045007} {\bibfield  {journal} {\bibinfo  {journal} {Physical Biology}\ }\textbf {\bibinfo {volume} {8}},\ \bibinfo {pages} {045007} (\bibinfo {year} {2011})}\BibitemShut {NoStop}%
\bibitem [{\citenamefont {Jia}\ \emph {et~al.}(2014)\citenamefont {Jia}, \citenamefont {Qian},\ and\ \citenamefont {Jiang}}]{Jia_overshoot_in_biologicalsystems}%
  \BibitemOpen
  \bibfield  {author} {\bibinfo {author} {\bibfnamefont {C.}~\bibnamefont {Jia}}, \bibinfo {author} {\bibfnamefont {M.}~\bibnamefont {Qian}},\ and\ \bibinfo {author} {\bibfnamefont {D.}~\bibnamefont {Jiang}},\ }\bibfield  {title} {\bibinfo {title} {Overshoot in biological systems modelled by markov chains: a non-equilibrium dynamic phenomenon},\ }\href {https://doi.org/10.1049/iet-syb.2013.0050} {\bibfield  {journal} {\bibinfo  {journal} {IET Systems Biology}\ }\textbf {\bibinfo {volume} {8}},\ \bibinfo {pages} {138} (\bibinfo {year} {2014})}\BibitemShut {NoStop}%
\bibitem [{\citenamefont {Miyahara}\ \emph {et~al.}(2023)\citenamefont {Miyahara}, \citenamefont {Yoneki},\ and\ \citenamefont {Roychowdhury}}]{Miyahara_VicsekMeetsDBSCAN}%
  \BibitemOpen
  \bibfield  {author} {\bibinfo {author} {\bibfnamefont {H.}~\bibnamefont {Miyahara}}, \bibinfo {author} {\bibfnamefont {H.}~\bibnamefont {Yoneki}},\ and\ \bibinfo {author} {\bibfnamefont {V.}~\bibnamefont {Roychowdhury}},\ }\bibfield  {title} {\bibinfo {title} {Vicsek model meets dbscan: Cluster phases in the vicsek model},\ }\href {https://doi.org/10.48550/arXiv.2307.12538} {\bibfield  {journal} {\bibinfo  {journal} {arXiv}\ } (\bibinfo {year} {2023})},\ \Eprint {https://arxiv.org/abs/2307.12538} {arXiv:2307.12538 [cond-mat.stat-mech]} \BibitemShut {NoStop}%
\bibitem [{\citenamefont {Peruani}\ and\ \citenamefont {Bär}(2013)}]{Peruani_kineticmodelandscalingproperties}%
  \BibitemOpen
  \bibfield  {author} {\bibinfo {author} {\bibfnamefont {F.}~\bibnamefont {Peruani}}\ and\ \bibinfo {author} {\bibfnamefont {M.}~\bibnamefont {Bär}},\ }\bibfield  {title} {\bibinfo {title} {A kinetic model and scaling properties of non-equilibrium clustering of self-propelled particles},\ }\href {https://doi.org/10.1088/1367-2630/15/6/065009} {\bibfield  {journal} {\bibinfo  {journal} {New Journal of Physics}\ }\textbf {\bibinfo {volume} {15}},\ \bibinfo {pages} {065009} (\bibinfo {year} {2013})}\BibitemShut {NoStop}%
\bibitem [{\citenamefont {Peruani}\ \emph {et~al.}(2010)\citenamefont {Peruani}, \citenamefont {Schimansky-Geier},\ and\ \citenamefont {Bär}}]{peruani_clusterdynamics_2010}%
  \BibitemOpen
  \bibfield  {author} {\bibinfo {author} {\bibfnamefont {F.}~\bibnamefont {Peruani}}, \bibinfo {author} {\bibfnamefont {L.}~\bibnamefont {Schimansky-Geier}},\ and\ \bibinfo {author} {\bibfnamefont {M.}~\bibnamefont {Bär}},\ }\bibfield  {title} {\bibinfo {title} {Cluster dynamics and cluster size distributions in systems of self-propelled particles},\ }\href {https://doi.org/10.1140/epjst/e2010-01349-1} {\bibfield  {journal} {\bibinfo  {journal} {The European Physical Journal Special Topics}\ }\textbf {\bibinfo {volume} {191}},\ \bibinfo {pages} {173} (\bibinfo {year} {2010})}\BibitemShut {NoStop}%
\bibitem [{\citenamefont {Starruß}\ \emph {et~al.}(2012)\citenamefont {Starruß}, \citenamefont {Peruani}, \citenamefont {Jakovljevic}, \citenamefont {Søgaard-Andersen}, \citenamefont {Deutsch},\ and\ \citenamefont {Bär}}]{Starrus_clusterdynamics_with_omega}%
  \BibitemOpen
  \bibfield  {author} {\bibinfo {author} {\bibfnamefont {J.}~\bibnamefont {Starruß}}, \bibinfo {author} {\bibfnamefont {F.}~\bibnamefont {Peruani}}, \bibinfo {author} {\bibfnamefont {V.}~\bibnamefont {Jakovljevic}}, \bibinfo {author} {\bibfnamefont {L.}~\bibnamefont {Søgaard-Andersen}}, \bibinfo {author} {\bibfnamefont {A.}~\bibnamefont {Deutsch}},\ and\ \bibinfo {author} {\bibfnamefont {M.}~\bibnamefont {Bär}},\ }\bibfield  {title} {\bibinfo {title} {Pattern-formation mechanisms in motility mutants of myxococcus xanthus},\ }\href {https://doi.org/10.1098/rsfs.2012.0034} {\bibfield  {journal} {\bibinfo  {journal} {Interface focus}\ }\textbf {\bibinfo {volume} {2}},\ \bibinfo {pages} {774} (\bibinfo {year} {2012})}\BibitemShut {NoStop}%
\bibitem [{\citenamefont {Yllanes}\ \emph {et~al.}(2017)\citenamefont {Yllanes}, \citenamefont {Leoni},\ and\ \citenamefont {Marchetti}}]{Yllanes_howmanydissenters}%
  \BibitemOpen
  \bibfield  {author} {\bibinfo {author} {\bibfnamefont {D.}~\bibnamefont {Yllanes}}, \bibinfo {author} {\bibfnamefont {M.}~\bibnamefont {Leoni}},\ and\ \bibinfo {author} {\bibfnamefont {M.~C.}\ \bibnamefont {Marchetti}},\ }\bibfield  {title} {\bibinfo {title} {How many dissenters does it take to disorder a flock?},\ }\href {https://doi.org/10.1088/1367-2630/aa8ed7} {\bibfield  {journal} {\bibinfo  {journal} {New Journal of Physics}\ }\textbf {\bibinfo {volume} {19}},\ \bibinfo {pages} {103026} (\bibinfo {year} {2017})}\BibitemShut {NoStop}%
\bibitem [{\citenamefont {{Chepizhko, O.}}\ and\ \citenamefont {{Peruani, F.}}(2015)}]{Chepizho_activeparticles_inheterogenous_media}%
  \BibitemOpen
  \bibfield  {author} {\bibinfo {author} {\bibnamefont {{Chepizhko, O.}}}\ and\ \bibinfo {author} {\bibnamefont {{Peruani, F.}}},\ }\bibfield  {title} {\bibinfo {title} {Active particles in heterogeneous media display new physics},\ }\href {https://doi.org/10.1140/epjst/e2015-02460-5} {\bibfield  {journal} {\bibinfo  {journal} {The European Physical Journal Special Topics}\ }\textbf {\bibinfo {volume} {224}},\ \bibinfo {pages} {1287} (\bibinfo {year} {2015})}\BibitemShut {NoStop}%
\bibitem [{\citenamefont {Martinez}\ \emph {et~al.}(2018)\citenamefont {Martinez}, \citenamefont {Alarcón}, \citenamefont {Rodriguez}, \citenamefont {Aragones},\ and\ \citenamefont {Valeriani}}]{Martinez_CollectiveBehaviorwithandwithouObstacles}%
  \BibitemOpen
  \bibfield  {author} {\bibinfo {author} {\bibfnamefont {R.}~\bibnamefont {Martinez}}, \bibinfo {author} {\bibfnamefont {F.}~\bibnamefont {Alarcón}}, \bibinfo {author} {\bibfnamefont {D.~R.}\ \bibnamefont {Rodriguez}}, \bibinfo {author} {\bibfnamefont {J.~L.}\ \bibnamefont {Aragones}},\ and\ \bibinfo {author} {\bibfnamefont {C.}~\bibnamefont {Valeriani}},\ }\bibfield  {title} {\bibinfo {title} {Collective behavior of vicsek particles without and with obstacles},\ }\bibfield  {journal} {\bibinfo  {journal} {The European Physical Journal E}\ }\textbf {\bibinfo {volume} {41}},\ \href {https://doi.org/10.1140/epje/i2018-11706-8} {10.1140/epje/i2018-11706-8} (\bibinfo {year} {2018})\BibitemShut {NoStop}%
\bibitem [{\citenamefont {Roy}\ and\ \citenamefont {Mugler}(2021)}]{roy2021intermediate}%
  \BibitemOpen
  \bibfield  {author} {\bibinfo {author} {\bibfnamefont {U.}~\bibnamefont {Roy}}\ and\ \bibinfo {author} {\bibfnamefont {A.}~\bibnamefont {Mugler}},\ }\bibfield  {title} {\bibinfo {title} {Intermediate adhesion maximizes migration velocity of multicellular clusters},\ }\href@noop {} {\bibfield  {journal} {\bibinfo  {journal} {Physical Review E}\ }\textbf {\bibinfo {volume} {103}},\ \bibinfo {pages} {032410} (\bibinfo {year} {2021})}\BibitemShut {NoStop}%
\bibitem [{\citenamefont {Mukherjee}\ and\ \citenamefont {Levine}(2021)}]{Mukherjee_ClusterSizeDistribution}%
  \BibitemOpen
  \bibfield  {author} {\bibinfo {author} {\bibfnamefont {M.}~\bibnamefont {Mukherjee}}\ and\ \bibinfo {author} {\bibfnamefont {H.}~\bibnamefont {Levine}},\ }\bibfield  {title} {\bibinfo {title} {Cluster size distribution of cells disseminating from a primary tumor},\ }\href {https://doi.org/10.1371/journal.pcbi.1009011} {\bibfield  {journal} {\bibinfo  {journal} {PLoS Computational Biology}\ }\textbf {\bibinfo {volume} {17}},\ \bibinfo {pages} {1} (\bibinfo {year} {2021})}\BibitemShut {NoStop}%
\end{thebibliography}%

\end{document}